\documentclass[11pt]{article}
\usepackage{amssymb,amsthm,latexsym,authblk,amsmath,pstricks,hyperref,multirow}
\usepackage{graphicx,epstopdf}
\usepackage{psfrag}
\usepackage{color} 
\usepackage{subfig}
\usepackage{epsfig}
\usepackage{epstopdf} 
\newcommand{\ket}[1]{| #1 \rangle}
\newcommand{\bra}[1]{\langle #1 |}
\newcommand{\braket}[2]{\langle #1 | #2 \rangle}
\newcommand{\ketbra}[2]{| #1 \rangle \langle #2 |}
\newcommand{\re}{\mathrm{Re}\, }
\newcommand{\im}{\mathrm{Im}\,}
\newcommand{\dss}{\displaystyle}

\newcommand{\I}{{\rm{i}}}
\newcommand{\D}{{\rm{d}}}
\newcommand{\E}{{\rm{e}}}
\newcommand{\tr}{\operatorname{tr}}

\newcommand{\identity}{\hat{\mathbf{1}}}
\newcommand{\onehalf}{{\frac{1}{2}}}

\newcommand{\adj}{{\rm h.c.}}

\newcommand{\aop}{\hat{a}}

\newcommand{\bop}{\hat{b}}

\newcommand{\nopA}{\hat{n}^A}
\newcommand{\nopAzero}{\hat{n}_0^A}
\newcommand{\nopAone}{\hat{n}_1^A}

\newcommand{\nopAj}{\hat{n}_j^A}
\newcommand{\nopvA}{\underline{\hat{n}}_A}
\newcommand{\dnopA}{ \Delta \hat{n}^A_{01}}
\newcommand{\nopB}{\hat{n}^B}
\newcommand{\nopBzero}{\hat{n}_0^B}
\newcommand{\nopBone}{\hat{n}_1^B}
\newcommand{\nopBi}{\hat{n}_i^B}
\newcommand{\nopBj}{\hat{n}_j^B}
\newcommand{\nopvB}{\underline{\hat{n}}_B}
\newcommand{\dnopB}{ \Delta \hat{n}^B_{01} }

\newcommand{\nv}{{\underline{n}}}

\newcommand{\nBzero}{{n}_0^B}
\newcommand{\nBone}{{n}_1^B}

\newcommand{\nvA}{{\underline{n}_A}}
\newcommand{\nvB}{{\underline{n}_B}}
\newcommand{\nuv}{{\underline{\nu}}}

\newcommand{\mvA}{{\underline{m}_A}}

\newcommand{\mv}{{\underline{m}}}
\newcommand{\pv}{{\underline{p}}}
\newcommand{\nuvA}{{\underline{\nu}_A}}
\newcommand{\nuvB}{{\underline{\nu}_B}}

\newcommand{\NopB}{\hat{N}_B}

\newcommand{\NB}{{N_B}}
\newcommand{\Hop}{\hat{H}}
\newcommand{\HA}{\hat{H}_A}
\newcommand{\HB}{\hat{H}_B}
\newcommand{\Hphi}{\hat{H}_\phi}
\newcommand{\HAB}{\hat{H}_{AB}}
\newcommand{\Htot}{\hat{H}_{AB}}

\newcommand{\Eint}{E^{(0)}}

\newcommand{\EGSB}{E_{0,B}^{(0)}}

\newcommand{\EGSA}{E_{0,A}^{(0)}}
\newcommand{\Kop}{\hat{K}}
\newcommand{\KA}{\hat{K}_A}
\newcommand{\KB}{\hat{K}_B}
\newcommand{\Kphi}{\hat{K}_\phi}
\newcommand{\Wop}{\hat{W}^{(01)}}

\newcommand{\WB}{\hat{W}_B^{(01)}}

\newcommand{\IB}{\hat{I}_B}
\newcommand{\PiB}{\hat{\Pi}_\Delta}
\newcommand{\inter}{{\rm int}}
\newcommand{\MI}{{\rm MI}}
\newcommand{\SF}{{\rm SF}}
\newcommand{\Jc}{{\mathcal J}}
\newcommand{\alphaA}{\alpha_A}
\newcommand{\alphaB}{\alpha_B}
\newcommand{\lambdaA}{\lambda_A}
\newcommand{\lambdaB}{\lambda_B}
\newcommand{\nuA}{\nu_A}
\newcommand{\nuB}{\nu_B}

\newcommand{\be}{\begin{equation}}
\newcommand{\ee}{\end{equation}}
\newcommand{\ba}{\begin{eqnarray}}
\newcommand{\ea}{\end{eqnarray}}

\def\integer{{\mathbb{Z}}}
\def\pinteger{{\mathbb{N}}}

\def\real{{\mathbb{R}}}

\def\proba{{\rm I\kern -.18em P}}

\newcommand{\eg}{e.g.}
\newcommand{\ie}{i.e.}
\newcommand{\Max}{{\rm max}}
\newcommand{\Min}{{\rm min}}

\newcommand{\Aa}{{\cal A}}

\newcommand{\Jj}{{\cal J}}
\newcommand{\Kk}{{\cal K}}

\newcommand{\Oo}{{\cal O}}

\newcommand{\Ss}{{\cal S}}

\newcommand{\Vv}{{\cal V}}

\begin{document}

\title{Persistent currents  in Bose-Bose mixtures after an interspecies interaction quench}


\author[1,2]{D. Spehner}
\affil[1]{\normalsize Departamento de Ingenier\'{\i}a Matem\'atica, Universidad de Concepci\'on, Concepci\'on, Chile}
\affil[2]{\normalsize Univ. Grenoble Alpes, CNRS, Institut Fourier and LPMMC,  F-38000 Grenoble, France}
\author[3]{L. Morales-Molina}
\affil[3]{\normalsize Facultad de F\'isica, Pontificia Universidad Cat\'olica de Chile, Casilla 306, Santiago, Chile}
\author[3]{S. A. Reyes}

 \date{\today}

\maketitle

 \begin{abstract}
  We study the persistent currents and interspecies entanglement generation in
  a Bose-Bose mixture formed by two atomic gases (hereafter labelled by the letters $A$ and $B$)
  trapped in a one-dimensional ring lattice potential with an artificial gauge field
  after a sudden quench from zero to strong  interactions between the two gases.
  Assuming that the strength of these interactions is much larger than the single species  energies
  and that the gas $A$ is initially in the Mott-insulator regime,
  we show that the current of the gas $B$ is reduced with respect to its value prior the interaction quench.
Averaging fast oscillations out, the  relative decrease of this  current
is independent of the initial visibility and Peierls phase of the gas $B$ and behaves quadratically with
the visibility of the gas $A$.
The second R\'enyi entropy of the reduced state measuring the amount of entanglement between the two gases is found to scale linearly
with the {number of sites}  and to be proportional to the 
relative decrease of the current.
 \end{abstract}

 \section{Introduction.}

The manifestations of quantum coherence and entanglement in many-body systems is one of the most challenging
problems in condensed matter physics and quantum technology.
The recent experimental realizations with trapped ultracold atoms of analogs of electronic mesoscopic systems
such as superconducting quantum interference devices~\cite{Campbell2013,Ryu2013,Eckel2014}
has opened new perspectives in the study of  matter-wave interferences. Atomtronics focuses on
the design of such atomic quantum devices, characterized by tunable parameters and low decoherence, and their applications to fundamental research and
technology~\cite{Seaman2007,Amico_book_2016}. 

One of the striking manifestation of quantum interferences is the Aharonov-Bohm effect. 
In a ring pierced by a magnetic flux, this gives rise to persistent currents varying periodically with the flux, which were observed long ago
in superconductors and normal metals~\cite{Deaver1961,Byers1961,Onsager1961,Levy1993}. In atomic
Bose-Einstein condensates (BECs) trapped in a ring-shaped potential, an artificial gauge field can be induced 
by laser fields or by the Coriolis force in the presence of a rotating potential barrier~\cite{Dalibard2011}.  
This system provides a novel platform for studying persistent currents~\cite{Kagan2000,Amico2018,Cominotti2014,Cominotti2015}. It
encompasses a rich variety of quantum phenomena,
including the formation of
macroscopic superpositions of clockwise and anticlockwise flowing states at
specific values of the magnetic flux
when rotation invariance is broken by a localized potential barrier~\cite{Leggett1980,Hallwood2010,Solenov2010,Nunnenkamp2011,Schenke2011}.
The influence  of such a barrier  on the current amplitude has been studied in Refs.~\cite{Cominotti2014,Cominotti2015}
for all strengths of the repulsive atomic interactions and in Ref.~\cite{Naldesi2019} for attractive
interactions.
{  The time evolutions of the persistent current and of space correlations
  when the gas is either slowly driven across the Superfluid (SF) to Mott-Insulator (MI) transition or is subject to an interaction quench
  crossing this transition
  have been also investigated~\cite{Kohn19,Kollath07}.}

The quantum interference effects become more complex when two atomic species are involved.
It is known that mixtures of two condensates can lead to the generation of exotic phases and to
the formation of quantum droplets,
among others phenomena~\cite{Inguscio2002,Esslinger2006,Wieman2008,Catani2008,Thalhammer2008,Inguscio2011,Cornish2011,Salomon2014,Cabrera2017}.
 It has been conjectured that the phase coherences of the two atomic gases in the mixing process may play a crucial role in these emerging phenomena~\cite{Catani2008,Thalhammer2008}.
 A number of theoretical and experimental works have focused on quantum phases  in optical lattices, in particular because of the
  analogy with condensed matter systems~\cite{kuklov1,kuklov2,kuklov3,Menotti,Li1,Li2,LMM,Penna,Moreira}.
    For repulsive interspecies interactions, {  phase separation processes have been investigated thoroughly
      (see e.g.~\cite{Suthar15,Suthar17}  and references therein for spectral and density profile analyses and~\cite{Mistakidis_NJP_2018}
      for quench dynamics crossing the miscibility to immiscibility transition).}
    More recently, Bose-Bose mixtures consisting of a single atom  interacting with a gas of atoms of a different species have attracted a lot of
    attention due to the formation of polarons in such systems~\cite{Meinert_Sceince_2017,Grust_NJP_2017,Volomiev_PRA2017,Mistakidis_PRL2019,Theel_NJP_2020}.

    A particular emphasis in the study of multi-component quantum gases concerns out-of-equilibrium dynamics following a sudden quench.
   Such quench dynamics can be investigated
   experimentally thanks to the high level of control on the trapping potential and the ability to tune (and even change the sign of) the
   intra- and inter-species interactions using Feshbach resonances~\cite{meinert,modu,Polleti}. 
   On the theoretical side, special attention has been devoted to interaction quenches (inter-species interactions are suddenly switched on)~\cite{Campbell_PRA_2014,morales,Mistakidis_NJP_2018,Mistakidis_PRL2019}.
   In these works, the generation of entanglement in the quench dynamics and its role in the observed phenomena have been investigated.  

   {  In the different context of the quench dynamics of a single component gas trapped in an infinite 1D-lattice,
     it has been shown that the entanglement associated to a partition of the lattice into a finite block of size $\ell$ and its complement,
      quantified by the von Neumann entropy of the reduced state,
  saturates to a value proportional to $\ell$ after a linear growth with time~\cite{Calabrese05}.
  Such a behavior of the entanglement after the quench
  has been observed experimentally for small lattice sizes  in measurements of the $2$-R\'enyi entropy of entanglement~\cite{Greiner16}.  
}

Coming back to persistent currents in BECs,  
it is  natural to ask about the effect on the current of the first atomic species of the
presence of a second atomic gas  trapped in the same ring. 
In order to change  this current, 
the two quantum gases must interact and be entangled with each other. One may then wonder whether the variation
of the current provides in some way a measure of the amount of entanglement { between the two gases.
Another natural question concerns the dependence of the persistent current on the  phase coherence of the two gases before they start to interact.

In this article, we investigate these questions by considering 
 a binary mixture containing two bosonic atomic species $A$ and $B$ (which may either correspond to different types of atoms or to identical atoms in different internal states) 
 trapped in a 1D-ring lattice potential in the presence of an artificial gauge field.
 Assuming that the two species are 
initially decoupled and that the gas $A$ is in the MI regime, we calculate analytically and numerically the time evolution of the current of $B$-atoms and the generation of
entanglement between the $A$ and $B$-condensates after a sudden quench from zero to strong inter-species
interactions.
 We quantify the amount of entanglement using the Schmidt number~\cite{my_review}  shifted by one,
  ${\cal K}_{AB}= \tr_A [ ( \tr_B \ketbra{\psi_{AB}}{\psi_{AB}} )^2]^{-1}-1 $, where $\ket{\psi_{AB}}$ is the wavefunction of the binary mixture
  after the interaction quench.
For small ${\cal K}_{AB}$ this quantity is nearly equal to the 2-R\'enyi entropy of entanglement
$S_{AB}^{(2)} = \ln ( 1+ {\cal K}_{AB} )$~\cite{Horodecki_review}, which is 
a measure of entanglement having similar properties as the entanglement of formation (von Neumann entropy of the reduced state). It has been shown recently that
$S_{AB}^{(2)}$ can be determined
from statistical correlations between random measurements~\cite{Beenaker_PRL2012,Vermersch_PRA2019,Vermersch_Science2019}
{  (see also~\cite{Daley12} for a measurement protocol of the 
  2-R\'enyi entropy, which can be applied in our setup to determine $S_{AB}^{(2)}$ in view of the possibility to monitor
 different trapping potentials for each species)}.

Our main results can be described as follows.
We show that the $B$-current after the quench is reduced due to interactions with the gas $A$. 
Averaging out the fast oscillations with frequency equal to the inter-species interaction energy divided by the
Planck constant, we prove that the relative  variation $\langle {\cal J}_B\rangle$
of the $B$-current before and after the quench
and the shifted Schmidt number $\langle {\cal K}_{AB}\rangle$ 
are proportional to each other and follow some universal laws. In particular, $\langle{\cal J}_B\rangle$ is independent of the visibility of the gas $B$
before the quench,
being the same when the latter is in the MI or in the SF regimes, and does also not depend on the Peierls phase of the $B$-atoms.
On the other hand, $\langle{\cal J}_B\rangle$ behaves quadratically with the initial visibility of
the gas $A$. 
Furthermore, $\langle {\cal K}_{AB}\rangle $ { and $S_{AB}^{(2)}$ }  are {  linear in the number of lattice sites}
$L$  and $\langle {\cal K}_{AB}\rangle/L =\beta \langle{\cal J}_B\rangle$,
with a proportionality factor $\beta$ depending only on the filling factor $\nu_B$, the Peierls phase, and the initial visibility of the gas $B$ when this gas is in the MI regime, whereas when it is in the SF regime $\beta$ depends  on the filling factor $\nu_B$ only.
We show that the generation of entanglement in the quench dynamics
  and its universal relation with the reduction of the $B$-current comes from 
  particle-hole
  excitations in the gas $A$  which slow down the flow of $B$-atoms and are
  coupled to site-dependent wavefunctions of the gas $B$.   
We argue that these results could be used for determining the amount of entanglement between the two gases from quantities
(atomic current and visibility of interference fringes) that can
be measured experimentally (see e.g.~\cite{Dalibard2014, Campbell2014, Mathew_PRA2015, Amico2018} for the observation of the supercurrent of a single-component
  gas in  ring-shape trapping potentials).

  Let us comment on the methods used to obtain  the aforementioned results.
  Our analytical calculations rely on (i) a perturbative expansion of the initial ground state and (ii)
  small and intermediate time approximations for the time propagator,
valid when the inter-species interaction strength is much larger than the
tunneling and intra-species interaction energies.
We point out that the method employed to determine the Schmidt number in Sec.~\ref{sec-Schmidt_number_intermediate_times} is original and could be useful in other contexts.
  It consists in expressing $\Kk_{AB} (t)$  when the gas $A$ is in the MI regime
  in terms  of an effective propagator acting on the other gas (see Appendix~\ref{app-B});
  by diagonalizing perturbatively the Hamiltonian 
  in this propagator for large inter-species interactions, we are able to determine  $\Kk_{AB} (t)$ at times of the order of
  the inverse tunneling and intra-species interaction energies of the gas $B$. 
  Our numerical simulations, in turn, rely on exact diagonalization. 
Although they  are carried out for small lattice sizes,
  these simulations corroborate the analytical results which apply to much larger atom numbers and lattice sizes.
This implies that finite  size effects are not
relevant for the physical effects we are interested in, thus enabling their
applications to mesoscopic systems of arbitrary size.

The paper is organized as follows. We introduce our model of a Bose-Bose mixture in a $1$D-ring lattice
in Sect.~\ref{sec-model}. Section~\ref{sec-visibility} is devoted to a brief discussion on the
visibility of a single species before the interaction quench and its behavior as function of the Peierls phase for finite lattice sizes.
The calculation of the $B$-{ current} after the interaction quench is performed in Sec.~\ref{sec-supercurrent}. We determine in Sec.~\ref{sec-Schmidt_number_small_times} the entanglement generation between the two gases
and show there that the time-averaged { shifted} Schmidt number $\langle \Kk_{AB} \rangle$ is proportional to $L$ and to
$\langle \Jj_B\rangle$, assuming an averaging time  much larger than the inverse inter-species interaction energy
and much smaller than the inverse single species energies.
{ In Sec.~\ref{sec-superpositions} we show that quantum superpositions in the post-quench total wavefunction involving particle-hole excitations in the gas $A$ are at the origin of the entanglement.}  
In Sec.~\ref{sec-Schmidt_number_intermediate_times}, we evaluate the  Schmidt number at larger times and
show that its average $\langle \Kk_{AB} \rangle_t$ remains constants over a large time period, suggesting
a convergence  of  $\langle \Kk_{AB} \rangle_t$  at large times $t$ to universal values which are  determined analytically.
Concluding remarks are given in Sec.~\ref{sec-conclusion}. Some technical details
on the analytical calculations  are presented in the two appendices.

\section{Model} \label{sec-model}

Let us consider two atomic gases $A$ and $B$ trapped in the same 1D-ring lattice potential with $L$ sites.
We denote by $\aop_{j}^{\dagger}$, $\aop_{j}$, and $\nopA_j= \aop_j^\dagger \aop_j$  
the creation, annihilation, and number operators  at site $j$ for atoms of the gas $A$,
and  by  $\bop_j^\dagger$, $\bop_j$, and $\nopB_j$ the corresponding operators for the gas $B$.
The two gases are described by Bose-Hubbard Hamiltonians, e.g.  for the gas $A$ 
\begin{equation} \label{eq-BH_Hamiltonian}
  \HA =- J_A\sum_{j}  (\E^{\I \phi_A} \aop^\dagger_{j+1} \aop_j + {\rm h.c.} )
  +\frac{U_A}{2}\sum_{j} \nopA_{j}(\nopA_{j}-1) = \KA + \HA^{\rm int} \; ,
\end{equation}
where $J_A$ and $U_A$ are the tunneling and interaction energy strengths and
 $\phi_A$ is the Peierls phase associated to an artificial gauge field.  
The sums in (\ref{eq-BH_Hamiltonian}) run over all
lattice sites $j=0$, \ldots, $L-1$ and ``h.c.'' refers to the Hermitian conjugate operator.
To account for the periodic boundary condition the sites $L$ and $0$ are identified with each other.
The Hamiltonian
$\HB$ of the gas $B$ is given similarly in terms of $J_B$, $U_B$, and $\phi_B$.
We consider repulsive intra-species interactions, $U_A,U_B>0$, and assume fixed total atom numbers $N_A$, $N_B$ for each species with integer
filling factors $\nu_A= N_A/L$, $\nu_B=N_B/L$.
The two gases are initially uncorrelated and in their ground states (GSs) $\ket{\psi_A}$ and $\ket{\psi_B}$.
At time $t=0$, attractive interactions between the two species are suddenly switched on (see the left panel in Fig.~\ref{fig-0}).
The mixture subsequently evolves according to the
Hamiltonian ($V>0$)
\begin{equation} \label{eq-HAB_int}
\Htot   =  \HA\otimes \identity_{B}+\identity_{A} \otimes \HB + \HAB^{\inter} 
\quad , \quad 
\HAB^\inter   =  - V  \sum_{j}  \nopA_{j} \nopB_{j}\;.
\end{equation}
We assume that  inter-species interactions are much larger than the single species
energies and that the gas $A$ is initially in the MI regime, \ie,
\begin{equation} \label{eq-separation_of_timescales}
 V  \gg U_A,U_B,J_B\quad ,\quad \lambdaA=\frac{J_A}{U_A}\ll 1\,.
\end{equation}
    { The gas $B$ can be either in the MI or the SF regime; in fact both
      cases $\lambdaB \ll 1$ and $\lambdaB \gtrsim 1$ will be analyzed in what follows.}

We will study in this work the time evolutions of the persistent current of $B$-atoms
(called in what follows the $B$-current) and of the amount of $A$-$B$ entanglement after the interaction quench.
The $B$-current is given at time $t\geq 0$  by
$I_B (t) = \tr_B [ \hat{\rho}_B (t) \IB ] $ with  $\hat{\rho}_B(t) = \tr_A \ketbra{\psi_{AB} (t)}{\psi_{AB} (t)}$ the reduced density matrix of the gas $B$,
$\ket{\psi_{AB} (t)}$ the time-evolved state of the mixture with the Hamiltonian (\ref{eq-HAB_int}), and $\IB$
the current operator
\begin{equation}
  \IB =  \frac{1}{2 L J_B}\frac{\partial \HB}{\partial \phi_B} = \frac{1}{2\I L} \sum_{j}
  \big( \bop_{j+1}^\dagger \bop_j \E^{\I \phi_B} - {\rm h.c.} \big)
\end{equation}
(here $\hbar = 1$).
{ We will focus in the following on the relative  variation of $B$-current and its average in the time interval $[0,t ]$,
defined by}
\begin{equation} \label{eq-relative_current_reduction}
  \Jc_B (t) = \frac{I_B(0) - I_B (t)}{I_B(0)}\quad ,\quad 
  \langle{\Jj_B}\rangle_t = \frac{1}{t} \int_0^t \D t' \Jc_B (t') 
  \,.
\end{equation}

When the tunelling energy of the gas $A$ vanishes (i.e. when $J_A=0$), the $B$-current is { time-independent} and thus $\Jj_B(t)=0$. Actually,
then the initial state $\ket{\psi_A} \ket{\psi_B}$ of the mixture has $\nuA$ atoms $A$ per site and
the coupling $\HAB^\inter$ acts  on it as $-V \nu_A \NB$,
where the total number $\NB=\sum_j \nopB_j$ of $B$-atoms 
is a $c$-number; hence
$\ket{\psi_{AB} (t)}$ is equal to $\ket{\psi_A} \ket{\psi_B}$ at all times $t$ up to irrelevant phases.
In contrast, when $J_A>0$  the  initial GS of the gas $A$
has particle-hole excitations~\cite{Gerbier05}, thus the coupling entangles the two gases
and the $B$-current is modified, $\Jj_B(t)\not= 0$.

The time evolution of the relative $B$-current variation, obtained numerically from
exact diagonalizations of $\HA$, $\HB$, and $\HAB$, is shown in the right panel of Fig.~\ref{fig-0}.
{ We use a second-order Suzuki-Trotter decomposition method with time steps
as small as $\Delta t=0.002$.
We observe that $\Jc_B (t)$} presents fast oscillations of period $2\pi/V$ superimposed on a complex
pattern with oscillations with larger periods~\cite{spectroscopy}.
The time-averaged version of $\Jc_B(t)$ displays two plateaus: the first one at times
$V^{-1} \ll t \ll U_A^{-1}, U_B^{-1}, J_B^{-1}$, the second one at times
of the order of the single species inverse energies.
We will show analytically in Sec.~\ref{sec-supercurrent}
that the value of the first plateau does not depend on the initial state of the gas $B$.
{ Note that this universal value holds for the relative variation of the $B$-current defined in (\ref{eq-relative_current_reduction});
  in contrast, the $B$-current strongly depends on the initial state of the gas $B$, being much smaller when $B$ is in the MI regime than when it is
  in the SF regime.}


Before calculating the $B$-current, we study the visibility of a single atomic species 
in a ring lattice with a gauge field before the interaction quench.

\begin{figure}
\begin{center}
  \begin{minipage}{5cm}
    \includegraphics[width=4cm,height=4cm, angle = 0]{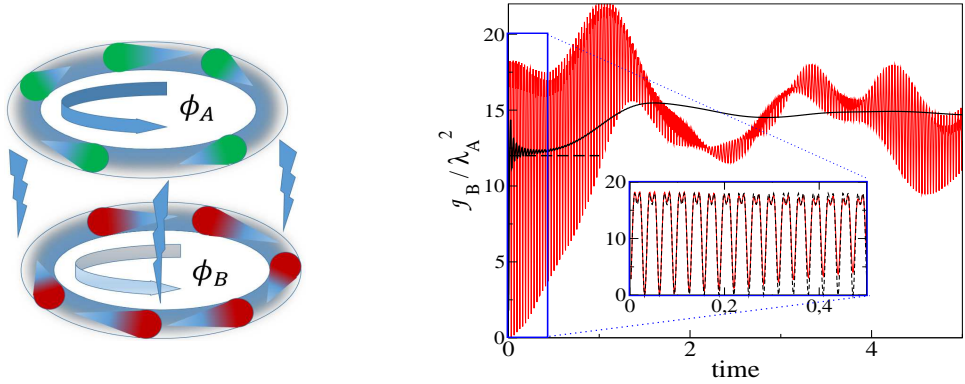}
  \end{minipage}
  \hspace{5mm}
  \begin{minipage}{7cm}
    \includegraphics[width=7cm,height=5cm, angle = 0]{Fig1b.eps}
\end{minipage}
\end{center}
\caption{
  {\bf Left panel:} Sketch of moving atoms on a 1D-ring lattice with two interacting  species $A$ and $B$ represented by green and red balls, in the presence of Peierls phases $\phi_A$ and $\phi_B$. 
{\bf Right panel:}
Time evolution after the interaction quench of the relative $B$-current variation
      ${\cal J}_B(t)$ (red curve) and its time-average (black curve), Eq.~(\ref{eq-relative_current_reduction}),
      divided by $\lambdaA^2$  for a lattice with $L=4$ sites,
      $N_A=N_B=4$ atoms of each species, $V=200 U_B$,
      $J_B=U_A=U_B$, $J_A=0.05 U_B$, $\phi_A=\phi_B=\pi/10$
      (from numerical calculations). Time is in units of $U_B^{-1}$.
      Dashed horizontal segment: universal value of Eq.~(\ref{eq-relation_current-variation_visib}).
      {\bf Inset:} amplification of the blue box shown in the figure with formula (\ref{eq-current_fluctuation_vs_time}) displayed in dashed line.     }
   \label{fig-0}
\end{figure}

\section{Single species visibility} \label{sec-visibility}

A good indicator of the degree of phase coherence  of a single species (say $B$) is  the visibility. The latter
is estimated experimentally by measuring the interference pattern after a free expansion of the gas. It is defined as
${\cal V}_B= ( S_{\max}-S_{\min})/(S_{\max}+S_{\min})$, 
$S_\Max$ and $S_\Min$ being the maximum and minimum of the momentum distribution
$S(q) = \sum_{i,j} \E^{\I q ( i-j)}\bra{\psi_B} \bop_i^\dagger \bop_j \ket{\psi_B}$.
 For $J_B=0$, the GS of the gas $B$ is the phase-incoherent
MI state
$\ket{\psi_B} = \ket{\psi_\MI}$  having $\nuB$ atoms ~{ per site~\cite{Gerbier05}}.
Then ${\cal V}_B=0$ (no interference fringes) and the $B$-current $\bra{\psi_B} \IB \ket{\psi_B}$ 
vanishes. Increasing the energy  ratio $\lambdaB=J_B/U_B$,
the visibility  and current increase, with ${\cal V}_B$ reaching
its  maximum ${\cal V}_B =1$ in the SF
limit $\lambda_B \gg 1$.

It has been shown both theoretically and experimentally~\cite{Gerbier05,Sengupta2005} that
the visibility of a single BEC trapped in an infinite gauge-free lattice potential behaves linearly with
$\lambda_B$  when $\lambda_B \lesssim 1$. However, we will need in the sequel the visibility
for finite lattice sizes in the presence of a gauge field and cannot rely on the results of these references.
In the MI regime $\lambdaB \ll 1$, one obtains by expanding 
perturbatively the GS $\ket{\psi_B}$ and the momentum distribution up to second order in  $\lambdaB$ that
(see Appendix~\ref{supplementary_material})
\begin{equation} \label{eq-visibility}
  {\cal V}_{B} =  4 (\nu_B+1) \lambda_B v_L ( \phi_B) \big[ 1 - (4 \nu_B+1) \lambdaB w_L (\phi_B)\big]
  +  \Oo (\lambdaB^3)
\end{equation}
with
\begin{eqnarray} \label{eq-def_v_L}
 v_L(\phi_B)  &\ = &
\begin{cases}
  \cos ( \phi_B - \ell \phi_0 )
   &  \text{if $L$ is even}
  \\[2mm]
  \frac{1}{2} \big( \cos  (\phi_B - \ell \phi_0 )  + \cos \big(  \frac{1+2k}{2}\phi_0  - \phi_B  \big) \big) 
   & \text{if $L$ is odd}
\end{cases}
\\ \label{eq-def_w_L}
 w_L(\phi_B)  & = & 
\begin{cases}
  0  
   &  \text{if $L$ is even}
  \\[2mm]
   - \cos  (\phi_B - \ell \phi_0 )  + \cos \big( \frac{1+2k}{2}\phi_0 - \phi_B \big) 
   & \text{if $L$ is odd,}
\end{cases}
\end{eqnarray}
where $\phi_0=2\pi/L$ is the lattice flux,
$\ell =  E [ \phi_B /\phi_0+ 1/2]$  the angular momentum of the SF state, and  $k= E [\phi_B /\phi_0]$ (here $E$ is the integer part).
Note that Gauge invariance implies that $ {\cal V}_{B}$ is periodic in the Peierls phase $\phi_B$ with period $\phi_0$.
In the limit $L \to \infty$ one finds that $v_L (\phi_B) , w_L (\phi_B) \to 1$  for any $\phi_B$,
so that $ {\cal V}_{B}$ becomes phase independent and
one recovers the result of Refs.~\cite{Gerbier05,Sengupta2005}.
It is easy to show from (\ref{eq-visibility})-(\ref{eq-def_w_L}) that for finite lattice sizes $L$, the visibility
$ {\cal V}_{B}$ reaches its minimum when  $\phi_B$ is equal to a half-integer value of $\phi_0$
(see Appendix~\ref{supplementary_material}).


In the opposite limit of weak interactions $U_B \ll J_B$ (SF regime  $\lambdaB \gg 1$),
when $\phi_B$ is not close to a half-integer value of $\phi_0$
the GS  of the Bose-Hubbard Hamiltonian can be approximated by the SF state (GS of the tunneling Hamiltonian).
Since the visibility in the latter state is equal to $1$ for all $\phi_B$'s, it follows that
${\cal V}_B =1 + \Oo (\lambda^{-1}_B )$.
The phases  $\phi_B =(\ell +1/2) \phi_0$, $\ell = 0,\pm1,\ldots$, are the points where
the parabola giving the energies of the SF states
with angular momenta $\ell$ and $\ell+1$ intersect. For such phases,
 quantum fluctuations of angular momentum produce
 fluctuations in the speed of rotation of the gas which are expected to blur the interference pattern,
 henceforth reducing the visibility as compared to its
 value  when $\phi_B$ is not close to a half-integer value of $\phi_0$.
 This is confirmed by numerical simulations for $L=3,4$, and $5$ (not shown here).

Comparing with our aforementioned result in the MI regime, 
we conjecture that {\it ${\cal V}_B$ is minimum at Peierls phases equal to half-integer values of $\phi_0$ 
  for any energy ratio $\lambda_B$}. This is supported by numerical calculations.
For instance, one observes in the left panel of Fig.~\ref{fig-1} that
${\cal V}_B$ presents a well pronounced minimum at $\phi_B=\phi_0/2=\pi/L$ when $L=3, 4$, and $5$ for
$\lambda_B=0.2$.
Note that the phase points $(\ell +1/2)\phi_0$ are singled out by the behavior of
the GS persistent current
$I_B (0)=\bra{\psi_B} \hat{I}_B \ket{\psi_B}$
   of the Bose gas as function of $\phi_B$. The latter is 
   periodic with period $\phi_0$ and drops rapidly and changes sign
   at these phase points,
   independently of the strength of the repulsive interactions
   (\ie,  of $\lambda_B$)~\cite{Leggett1980,Naldesi2019}.
   For reference, the behavior of $I_B(0)$ as function of $\phi_B$   
   is shown in the right panel in Fig.~\ref{fig-1}.

   Note that in Fig.~\ref{fig-1} the depth of the minima of ${\cal V}_B$
     become less pronounced as $L$ is increased,
 in agreement with the expectation that the visibility is phase independent for an infinite ring.
 Let us stress that the numerical simulations carried out in this work
consider small lattice sizes. This enables us to observe the dependence of
 physical quantities like the visibility on the Peierls phase,
whose effect vanishes at large sizes.

\begin{figure}
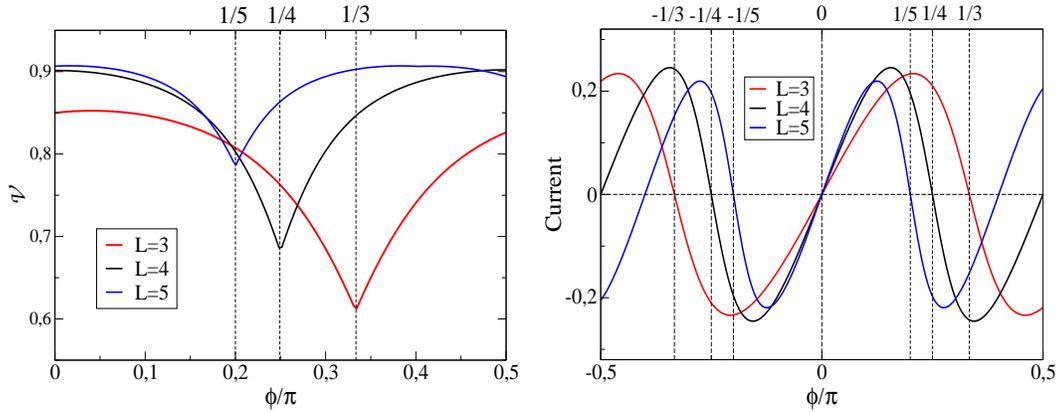

\begin{center}
  \begin{minipage}{7cm}
       \includegraphics[width=6.8cm,height=5.4cm,angle=0]{Fig2a.eps}
  \end{minipage}
\begin{minipage}{7cm}
       \includegraphics[width=6.8cm,height=5.4cm,angle=0]{Fig2b.eps}
\end{minipage}
\end{center}
\caption{
  {\bf Left panel:} 
      Visibility of a Bose gas with a single species $B$ trapped in a  1D-ring lattice potential
      as function of the Peierls phase $\phi_B$ for $\lambda_B=J_B /U_B =0.2$ and
      $N_B=L=3,4$, and $5 $ (from numerical calculations). 
 {\bf Right panel:}  $B$-current $I_B(0)$ (in arbitrary units) of the gas $B$ prior the interaction quench
as function of $\phi_B$
      for  the same parameters. The value of $\lambda_B$ is close to the transition between
      the MI and SF regimes.
}
\label{fig-1}
\end{figure}

\begin{figure}
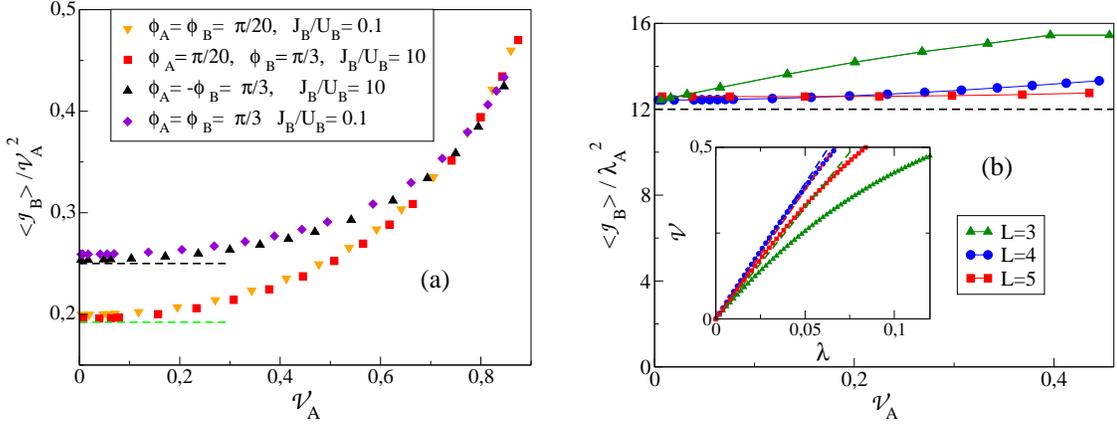

  \begin{center}
    \includegraphics[width=7cm, angle = 0]{Fig3a.eps}
    \hspace{0.5cm}
    \includegraphics[width=7cm,angle=0]{Fig3b.eps}
    \end{center}
  \caption{{\bf (a)} Time averaged current variation $\langle{\Jc_B}\rangle_t$ divided by the square visibility
    ${\cal V}_A^2$ vs.   $\Vv_A$ for different values of $\lambda_B=J_B/U_B$ and Peierls phases, from numerical calculations
     with  $L=N_A=N_B=4$. The average is up to time $t = 0.3/U_B$.
    { Triangles downward:} $V=100 U_B$,
    $U_A=U_B$, $J_B=0.1 U_B$, $\phi_A=\phi_B=\pi/20$.
    {Squares:} $V=1000 U_B$, $U_A=J_B=10 U_B$, $\phi_A=\pi/20$, $\phi_B=\pi/3$.
    {Triangles upward:} $V=1000 U_B$, $U_A=J_B=10 U_B$, $\phi_A=-\phi_B=\pi/3$.
    {Diamonds:} $V=100 U_B$, $U_A=U_B$, $J_B=0.1 U_B$, $\phi_A=\phi_B=\pi/3$. In all cases $J_A$ is varied in order
    to change ${\cal V}_A^2$.
    The horizontal segments correspond to the values in the RHS of Eq.(\ref{eq-relation_current-variation_visib}).
    {\bf (b)} $\langle{\Jc_B}\rangle_t$ divided by $\lambdaA^2$ for  $V=100 U_B$, $U_A=U_B$, $J_B=0.1 U_B$, $J_A$ variable, $\phi_A=\phi_B=\pi/20$, and different site numbers  $L=N_A=N_B=3, 4$ and $5$. The horizontal dashed line corresponds to the first equality in Eq.(\ref{eq-relation_current-variation_visib}). 
    {\bf Inset:}~single species visibility vs $\lambda=J/U$ for the same values of $L$ and $N$, with dashed lines
    showing the linear term in Eq.~(\ref{eq-visibility}).\\
 }
\label{fig-2}
\end{figure}

\section{The $B$-current} \label{sec-supercurrent}

We can calculate  the $B$-current  at times $t \ll U_A^{-1}, U_B^{-1}, J_B^{-1}$ as follows.
Using the invariance of the reduced density matrix $\hat{\rho}_B (t)$ with respect to translations by one lattice site and introducing
the eigenstates $\ket{\nvB}=\ket{n^B_0,\ldots,n_{L-1}^B}$ of
$\nopvB=(\nopB_0$, \ldots, $\nopB_{L-1})$ (Fock states),  one has
\begin{equation} \label{eq-general_expression_B_current}
  I_B (t) =  \im \sum_{\nvB} \! \sqrt{\nBzero ( \nBone +1)}
   \, \E^{\I \phi_B} \bra{\nvB} \hat{\rho}_B(t) \ket{\nvB + 1_1 -1_0}  \;, 
\end{equation}
where $1_i$ denotes the vector with components $\delta_{ij}$, $j=0, \ldots,L-1$, and the sum runs over all $\nvB \in \pinteger^L$, 
$\sum_j n_j^B = \NB$.
Due to our small time hypothesis,
hopping and intra-species interactions can be neglected in the Hamiltonian (\ref{eq-HAB_int})
and the dynamics  is solely governed by inter-species interactions. Thus
$\ket{\psi_{AB}(t)} = \E^{\I t V \nopvA \cdot \nopvB} \ket{\psi_A}\ket{\psi_B}$.
Plugging this expression into~(\ref{eq-general_expression_B_current}), the current can be cast as 
\begin{equation}
I_B (t) =
I_B(0)  \langle \E^{- \I t V \dnopA} \rangle_{\psi_A}\quad , \quad \dnopA= \nopAone - \nopAzero\;,
\end{equation}
where $I_B(0)$ is the current in the GS of $\HB$ (obtained by substituting $\hat{\rho}_B(t)$ by
$\ketbra{\psi_B}{\psi_B}$ in (\ref{eq-general_expression_B_current})), $\langle \cdot \rangle_{\psi_A} = \bra{\psi_A} \cdot \ket{\psi_A}$
is the quantum expectation in $\ket{\psi_A}$,
and we have used that $\langle \E^{- \I t V \dnopA} \rangle_{\psi_A}$ is real, as will be shown below. 
The relative current variation reads 
\begin{equation} \label{eq-relative_current_reduction_small_times}
  \Jc_B (t) = 1 - \big\langle \E^{- \I t V \dnopA} \big\rangle_{\psi_A}
  \;\; ,  \;\; t \ll U_A^{-1}, U_B^{-1}, J_B^{-1}\, .
\end{equation}  
It follows from the Cauchy-Schwarz inequality 
that $\Jc_B(t) \geq 0$. Hence { the effect of the inter-species coupling at time $t>0$ is} to reduce the currents of each species. Furthermore,
{ the relative current variation} $\Jc_B (t)$ does not depend on the initial GS of the gas $B$, being the same when $B$ is initially in the SF or MI regimes, and is also independent of the Peierls phase $\phi_B$ of the gas $B$.
  This independence of the relative variation of $B$-current after the interaction quench   from $\phi_B$ and $\lambda_B$
  contrasts with the behavior of the $B$-current  $I_B(0)$ before the quench. The latter oscillates with $\phi_B$
  (see the right panel in Fig.~\ref{fig-1}) with an amplitude which strongly depends on
  the visibility $\Vv_B$, \ie, on $\lambda_B$ (see the discussion at the beginning of Sec.~\ref{sec-visibility}). 
  The independence of  $\Jj_B (t)$ from $\lambda_B$ and $\phi_B$ is confirmed by
  the numerical results displayed in  
  Fig.~\ref{fig-2}(a). Indeed, one sees in this figure that the time-averaged values $\langle \Jj_B\rangle_t$
  of $\Jj_B (t)$ for $\lambda_B=10$, $\phi_B=\pi/3$ and for $\lambda_B=0.1$, $\phi_B=\pi/20$ are very close from each other,
  for all values of the visibility $\Vv_A$  of the gas $A$. As we shall explain below,
  the observed dependence of $\langle{\Jj_B}\rangle_t/ \Vv_A$ 
  on the Peierls phase $\phi_A$ of the gas $A$ for small $\Vv_A$
  originates from the variation of $\Vv_A$ with $\phi_A$ predicted by Eq. (\ref{eq-visibility}); this is thus  a
  finite-size effect.

We next calculate the $A$-expectation in (\ref{eq-relative_current_reduction_small_times}), assuming that
the gas $A$ is in the MI regime ($\lambda_A \ll 1$), and show in the way
that it is real up to second order in $\lambdaA$. (Let us note that although this is true for all values of $\lambdaA$ when $\phi_A=0$ by symmetry of 
  $\HA$ under $j\mapsto L-j \mod(L)$, this is not the case for  $\phi_A \not= 0$ and large $\lambdaA$.)
The GS of $\HA$  can be evaluated by treating
the tunneling Hamiltonian
$\KA$ in (\ref{eq-BH_Hamiltonian}) perturbatively.
The unperturbed Hamiltonian $\HA^\inter$  has eigenvectors $\ket{\nvA}$, eigenenergies $E_A^{(0)} (\nvA)$,
and GS $\ket{\psi_\MI}=\ket{\nuvA}$ with $\nuvA=(\nu_A,\ldots, \nu_A)$.
Noting that the energy to create a particle-hole excitation
is $E_A^{(0)}(\nuvA \pm 1_{i+1} \mp 1_i) - E_{A}^{(0)} ( \nuvA) = U_A$, one finds
\begin{equation} \label{eq-pertubrative_expansion_GS}
  \ket{\psi_A} = \ket{\psi_\MI} + \lambdaA \ket{\psi^{(1)}_A} + \lambdaA^2 \ket{\psi^{(2)}_A} + \Oo ( \lambdaA^3)\;,
\end{equation}
where the first-order correction is 
\begin{equation} \label{eq-1st_order_corr_GS}
\ket{\psi^{(1)}_A} = - \frac{1}{J_A} \KA \ket{\psi_\MI} =  \sum_j \big( \E^{\I \phi_A} \aop_{j+1}^\dagger \aop_j  + \adj \bigr) \ket{\psi_\MI}
\end{equation}
and the second-order correction satisfies (see appendix~\ref{supplementary_material})   
\begin{equation}  \label{eq-2nd_order_correc_GS}
  \braket{\psi_\MI}{\psi^{(2)}_A}   = -  \alpha_A L\;\text{ with }\; \alpha_A= \nu_A(\nu_A+1)\,.
\end{equation}  
A simple calculation yields
{
  \begin{equation}
 \big\langle \E^{- \I t V \dnopA} \big\rangle_{\psi_A^{(1)}} \equiv \bra{\psi_A^{(1)}} \E^{- \I t V \dnopA} \ket{\psi_A^{(1)}}     
 = 2 \alpha_A \sum_j \cos( t V  \Delta_{01}^{j, j+1} )\;,
  \end{equation}
  where  $\pm \Delta_{01}^{j, j+1}$ stands for the eigenvalue of $ \dnopA$ in the
  particle-hole excitation state
  \begin{equation} \label{eq-particle-hole_state}
    \ket{\varphi_{A,j,\pm}}= \ket{ \nuv_A \pm 1_{j+1} \mp 1_j}\;.
  \end{equation}
  Since 
  $\Delta_{01}^{j, j+1}= \delta_{j+1,1}-\delta_{j,1}- \delta_{j+1,0}+\delta_{j,0}$ equals $2$ if $j=0$, $\Delta_{01}^{j, j+1}=-1$ if $j=1$ or $L-1$,
  and $\Delta_{01}^{j, j+1}=0$ otherwise,
  one finds
}  
\begin{equation} \label{eq-expectation_in_first_order_WF}
  \big\langle \E^{- \I t V \dnopA} \big\rangle_{\psi_A^{(1)}} = 
  2 \alpha_A [ L -  3 + 2 \cos(tV ) + \cos (2 tV ))]\,.
\end{equation}
Using  (\ref{eq-relative_current_reduction_small_times}),
(\ref{eq-2nd_order_correc_GS}), 
(\ref{eq-expectation_in_first_order_WF}), $\E^{-\I t V \dnopA} \ket{\psi_\MI} = \ket{\psi_\MI}$, and $\braket{\psi^{(1)}_A}{\psi_\MI}=0$, this gives
\begin{equation} \label{eq-current_fluctuation_vs_time}
\Jc_B (t)=  2 \alpha_A \lambdaA^2 [ 3 - 2 \cos(tV) - \cos ( 2 t V )] + \Oo(\lambdaA^3)\,. 
\end{equation}
A good agreement between Eq. (\ref{eq-current_fluctuation_vs_time}) and the numerical result
at times $t \ll 1/U_A,1/U_B,1/J_B$ is seen in  Fig.~\ref{fig-0} (inset, right panel).
The cosines disappear upon averaging up to a time  $t$ much larger than $1/V$, with $t \ll 1/U_A,1/U_B,1/J_B$. Plugging
the expression of the visibility ${\cal V}_A$ analogous to (\ref{eq-visibility}), one gets
\begin{equation} \label{eq-relation_current-variation_visib}
  \langle{\Jj_B}\rangle_t  = 6 \alpha_A \lambdaA^2 =   \frac{3 \nuA}{8(\nuA+1)} \frac{{\cal V}_A^2}{v^2_L (\phi_A )}
\end{equation}  
with errors of order $\lambdaA^3$.
We see that {\it $\langle{\Jj_B}\rangle_t$ is independent of the initial coherence, number of atoms,
  and Peierls phase of the gas $B$ and of the size $L$ of the ring}. Moreover, it  depends quadratically on the initial visibility of the gas $A$. 
Recalling  that  $v_L (\phi_A ) \to 1$ for $L \to \infty$, the dependence on the Peierls phase $\phi_A$ of the 
  expression in the last member of Eq.~(\ref{eq-relation_current-variation_visib}) disappears in the infinite size limit.

Comparing Eq.~(\ref{eq-relation_current-variation_visib}) with the numerical results
shown in Fig.~\ref{fig-2}(a), which displays $\langle {\Jj_B}\rangle_t/ {\cal V}_A^2$  
  as function of  ${\cal V}_A$
  for two different values of $\lambdaB$ corresponding to the gas $B$ being initially in the MI and SF regimes and for different phases,
{ a good agreement is observed  when ${\cal V}_A \ll 1$}, the values calculated numerically
being about $5\%$ above the
prediction of Eq.~(\ref{eq-relation_current-variation_visib}).
Clear deviations from Eq. (\ref{eq-relation_current-variation_visib}) show up for
${\cal V}_A \gtrsim 0.2$, as expected from the fact that this equation is valid for $\lambda_A \ll 1$ only.
However, Fig.~\ref{fig-2}(b) shows that for such visibilities these deviations become smaller
by increasing the lattice size.

  { The physical origin of the reduction of the $B$-current when the two  gases are coupled at time $t>0$
    can be understood by looking at the effect on the $B$-atoms  of particle-hole excitations in the gas $A$
    (Note that the MI state of the gas $A$ has no effect on the $B$-current since the attractive potential
    produced on each $B$-atom by its coupling with the gas $A$ in the MI state is site-independent and equal to $-V \nuA$;
    in other words, if the two gases are in the state
    $\ket{\psi_\MI} \ket{\psi_B}$ then their coupling energy 
    is independent  of the  distribution of the $B$-atoms on the lattice and is equal to
    $-V \nu_A N_B$,  as  noted in Sec.~\ref{sec-model}.)
    Consider e.g. a gas $A$ in the particle-hole state $\ket{\varphi_{A,j,+}}$. Then the transfer of a $B$-atom  from site $j+1$
    to site $j$ has a coupling energy cost of $ 2 V$ (in fact, the coupling energy increases by $V (\nu_A+1)$  when removing the $B$-atom from
    site $j+1$ and decreases by
    $V(\nu_A-1)$ when adding it on site $j$). Similarly, if the gas $A$ is in the particle-hole state  $\ket{\varphi_{A,j,-}}$, the transfer
    of a $B$-atom  from site $j$ to site $j+1$ has an energy cost of $2V$. We can infer that
    particle-hole excitations in the gas $A$ slow down the flow of $B$-atoms
    in the ring lattice. We shall come back to this effect in Sect.~\ref{sec-superpositions} after having studied the entanglement generation
    between the two gases.
  }

\section{Interspecies entanglement and its relation with the $B$-current}
\label{sec-Schmidt_number_small_times}

The reduction of current  is due to quantum correlations between the $A$- and $B$-atoms induced by
the interspecies coupling.
We estimate the amount of entanglement between the two gases 
using the shifted Schmidt number
\begin{equation} \label{eq-def_Schmidt_number}
  \Kk_{AB} (t)=  \big( \tr_A[\hat{\rho}_A^2 (t)] \big)^{-1} - 1
  = \big( \tr_B[\hat{\rho}_B^2 (t)] \big)^{-1} - 1 \;,
\end{equation}
where  $\hat{\rho}_A(t) = \tr_{B} \ketbra{\psi_{AB} (t)}{\psi_{AB} (t)}$ is the reduced density matrix of the gas $A$ at time $t$.
The shift by one in our definition of the Schmidt number, as compared with the usual definition, makes sure that
 $\Kk_{AB}(t)=0$ when the two gases are unentangled. Note that $ \Kk_{AB} (t)$ is symmetric under the exchange of $A$ and $B$.

Assuming as before that  the gas $A$ is in the MI regime and
considering times $t$ much smaller than  $U_A^{-1}$, a simple calculation using the perturbative
expansion (\ref{eq-pertubrative_expansion_GS}) and  neglecting the hopping of atoms $A$ in the dynamics yields
(see in Appendix~\ref{app-B})
\begin{equation} \label{eq-Schmidt_number_general_expression} 
\Kk_{AB} (t) = 4 L \alpha_A  \lambdaA^2 \Big( 1 - \big| \big\langle \E^{-\I t \WB} \big\rangle_{\psi_B} \big|^2 \Big) + \Oo (\lambdaA^3)
\end{equation}  
with the Hamiltonian
\begin{equation} \label{eq-def_WB}
  \WB = \HB - V \dnopB
  \quad , \quad \dnopB = \nopB_1-\nopB_0 \;.
\end{equation}
In this section, like for the calculation of the $B$-current,  we restrict our analysis
to times $t$ much smaller than all the inverse single species energies. 
We refer to this condition as defining the {\it short time regime}. The behavior of the Schmidt number at larger times
will be discussed in Sec.~\ref{sec-Schmidt_number_intermediate_times}. By applying (\ref{eq-Schmidt_number_general_expression}), we first
determine $\Kk_{AB} (t)$ in the short time regime in two limits: $\lambda_B \ll 1$ (gas $B$ in the MI regime) in Subsect.~\ref{eq-entanglement_B_MI_regime}
and $\lambda_B \gg 1$ (gas $B$ in the SF regime) in Subsect.~\ref{eq-entanglement_B_SF_regime}. We then 
discuss the general case in Subsect.~\ref{entanglement_arbitary_lambda_B}.

\subsection{Gases $A$ and $B$ in the MI regime} \label{eq-entanglement_B_MI_regime}

Let us  first assume that 
both gases are in the MI regime ($\lambdaA,\lambdaB \ll 1$).
For times $t \ll U_A^{-1}, U_B^{-1}$, the Hamiltonian $\HB$ 
can be replaced by $\HB^\inter$ in Eq.~(\ref{eq-Schmidt_number_general_expression}) and thus
be dropped out
(recall that $[\HB^\inter, \dnopB]=0$ and $\HB \ket{\psi_B}=E_{\rm GS}^B \ket{\psi_B}$). The quantum expectation in this equation
becomes the same as in the calculation of the $B$-current, replacing $A$ by $B$. This yields
\begin{equation} \label{eq-Schmidt_number_small_lambdaA_lambdaB}
  \Kk_{A B} (t) = 16 L \alpha_A \alpha_B \lambda_A^2 \lambda_B^2 \big[ 3 - 2 \cos(tV)-\cos (2 t V) + \Oo (\lambda_A) + \Oo ( \lambda_B) \big]
 \end{equation}
%
with $\alpha_B=\nu_B ( \nu_B+ 1)$. Averaging up to time $t$ with $V^{-1} \ll t \ll U_A^{-1},U_B^{-1}$
one gets
\begin{equation} \label{eq-average_Schmidt_number_small_lambdaA_lambdaB}
  \langle {\Kk_{A B}} \rangle_t =  48 L \alpha_A \alpha_B \lambdaA^2 \lambdaB^2 = \frac{3L}{16} \frac{\nuA  \nuB}{(\nuA+1)(\nuB+1)}
  \frac{\Vv_A^2 \,\Vv_B^2}{v_L(\phi_A)^2 v_L (\phi_B)^2} \;,
\end{equation}
where we have used  in the last equality the linear approximation of the visibility, see (\ref{eq-visibility}).

In Eqs.~(\ref{eq-Schmidt_number_small_lambdaA_lambdaB}) and~(\ref{eq-average_Schmidt_number_small_lambdaA_lambdaB}) we can observe   that the shifted Schmidt number {  is proportional to} the system size $L$. Note that the $2$-R\'enyi entropy of entanglement, which is a meaningful
  entanglement measure analogous to the entanglement of formation~\cite{Horodecki_review}, is related to
  the Schmidt number defined in \eqref{eq-def_Schmidt_number} by
\begin{equation}  
  S^{(2)}_{AB} ( t)  = \ln (  {\Kk_{A B}}(t) +1) \simeq   {\Kk_{A B}}(t)\;,
\end{equation}
where the second equality holds for $ \Kk_{A B} (t) \ll 1$. The latter condition is satisfied  for not too large $L$'s in Eq.~(\ref{eq-average_Schmidt_number_small_lambdaA_lambdaB}) since we assume $\lambda_A,\lambda_B \ll 1$.  
Therefore, {  the entanglement entropy after the quench scales linearly with the system size $L$.}

Furthermore, by comparing (\ref{eq-relation_current-variation_visib}) and
(\ref{eq-Schmidt_number_small_lambdaA_lambdaB}) one finds that
${\Kk_{A B}} (t)$ { (and thus $ S^{(2)}_{AB} ( t) $)}  is  proportional to the relative $B$-current reduction, 
\begin{equation} \label{eq-proportionality_Schmidt_number_relative_current}
  S^{(2)}_{AB} ( t) /L=  {\Kk}_{A B}(t) /L = \beta_{\nuB,\lambdaB} \Jc_B (t)
 \end{equation}
with a proportionality factor $\beta_{\nuB,\lambdaB} =8 \alphaB \lambdaB^2$ depending on the filling factor $\nuB$ and the energy ratio $\lambdaB$
of the $B$-gas only. Using (\ref{eq-visibility}) again, one has  
\begin{equation} \label{eq-prop_factor_lambda_B_small}
  \beta_{\nuB,\lambdaB}  =8 \alpha_B \lambdaB^2=\frac{\nuB}{2\nuB+2} \frac{\Vv_B^2}{v^2_L(\phi_B)}\;,
\end{equation}
showing that ${\Kk}_{A B}(t) /L$ is proportional to $ \Vv_B^2  {\Jc_B} (t)$.

{ It is worth noting that the time-averaged Schmidt number per site has a universal
  value $48 \alpha_A \alpha_B \lambda_A^2 \lambda_B^2$, see (\ref{eq-average_Schmidt_number_small_lambdaA_lambdaB}),
  which is independent of the Peierls phases and, more remarkably, of the strength $V$ of the inter-species interactions.
  Moreover,  the amplitude of the time oscillations of ${\Kk}_{A B}(t)$ has also a universal character, see
  (\ref{eq-Schmidt_number_small_lambdaA_lambdaB}); in particular it does not depend on
  $V$ which only sets in the period of  oscillations.}

Since both the $B$-current and the visibilities   are measurable quantities,
Eqs.~(\ref{eq-average_Schmidt_number_small_lambdaA_lambdaB}) and (\ref{eq-proportionality_Schmidt_number_relative_current}) 
provide a way to evaluate the amount of $A$-$B$ entanglement.
This can be done either by measuring  the interference patterns
of the two gases prior to the interaction quench (Eq.~(\ref{eq-average_Schmidt_number_small_lambdaA_lambdaB}))
or by performing measurements prior and after the quench on the gas $B$ only (Eq.~ (\ref{eq-proportionality_Schmidt_number_relative_current})).
{ Note that the usual technique to estimate entanglement  experimentally in quantum many-body systems relies on quantum state tomography
  (see~\cite{Treutlein_NJP2011,Lanyon_Nature2017} and references therein) and can be applied in practice to small systems only
  (see also alternative techniques proposed in~\cite{Mistakidis_NJP_2018,Vermersch_PRA2019,Vermersch_Science2019,Daley12}).
  In contrast,  estimating entanglement from observations of the interference patterns and measurement of the atomic
  current can be done without any restriction on the system size.}

The numerical results displayed in Fig.~\ref{Fig-3} show good agreement with
Eqs.  (\ref{eq-average_Schmidt_number_small_lambdaA_lambdaB}) and
(\ref{eq-proportionality_Schmidt_number_relative_current}) for values of $\lambdaA$ corresponding to ${\cal V}_A \lesssim 0.2$.
For higher visibilities ${\cal V}_A$, it is seen in the left panel that $\langle {\Kk}_{AB} \rangle_t$ depends on the Peierls phases.
Note that for large $L$, (\ref{eq-proportionality_Schmidt_number_relative_current} and (\ref{eq-prop_factor_lambda_B_small}) give
$\langle \Kk_{AB}\rangle_t/L = ( \nu_B/(2\nu_B+2) ) \Vv_B^2 \langle \Jj_B\rangle_t$, hence the slopes of the
straight lines in the right panel become $\phi_A$-independent and equal to $1/4$.

\begin{figure}
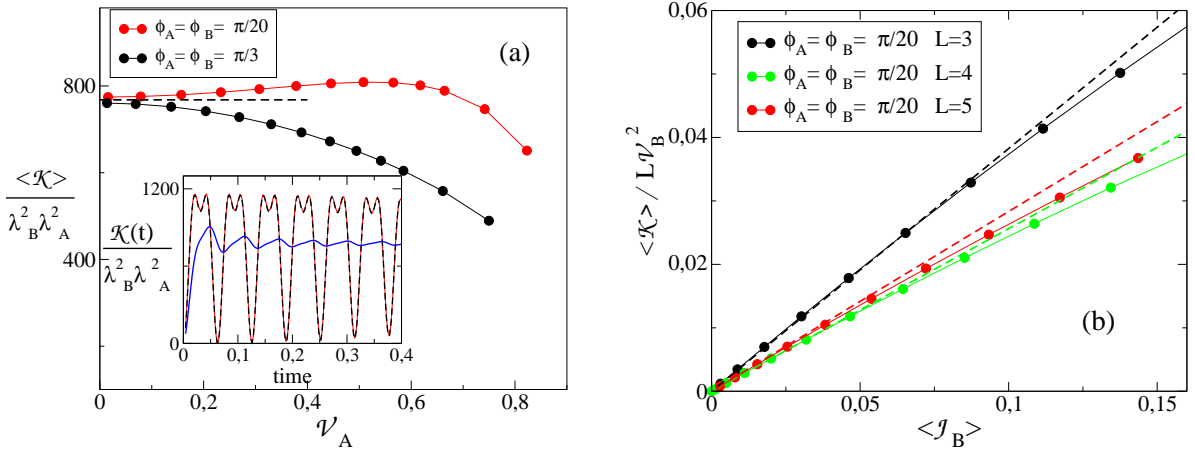

    \begin{center}
      \includegraphics[width=7.5cm, angle = 0]{Fig4a.eps}
      \hspace{0.5cm}
      \includegraphics[width=7.5cm, angle = 0]{Fig4b.eps}
    \end{center}
    \caption{{\bf (a):}
      Time averaged Schmidt number $\langle \Kk_{AB} \rangle_t$
      divided by $\lambda_A^2 \lambda_B^2$  vs.
      initial visibility of the gas $A$ for a gas $B$ in the MI regime. The 
      parameters are $L=N_A=N_B=4$, $V=100 U_B$, $J_B=0.02 U_B$, $U_A=0.1 U_B$, $J_A$ variable,
      $\phi_A=\phi_B=\pi/20$ (red dots) and
      $\pi/3$ (black dots), averaging time $t=0.3/U_B$. Horizontal dashed segment: value of
      Eq. (\ref{eq-average_Schmidt_number_small_lambdaA_lambdaB}).
      {\bf Inset:}~Schmidt number vs time (in units of $U_B^{-1}$) for the same parameters and $J_A=0.0002 U_B$.
      Red and blue line: $\Kk_{AB}(t)$ and $\langle \Kk_{AB}\rangle_t$ from numerical calculations.
      Dashed black lines: formula (\ref{eq-time_evolution_of_K}).  
      {\bf (b):} $\langle \Kk_{AB} \rangle_t/L$ divided by the squared visibility of the gas $B$ vs.
      relative current reduction $\langle {\cal J}_B \rangle_t$, for the same energy parameters, 
      $\phi_A=\phi_B=\pi/20$, and $L=N_A=N_B=3,4$ and $5$ (black, green, and red points). 
      Dashed lines: values from Eqs. (\ref{eq-proportionality_Schmidt_number_relative_current}) and
      (\ref{eq-prop_factor_lambda_B_small}).\\
     }
  \label{Fig-3}
\end{figure}

\begin{center}
\begin{table}[t]
  \centering
  \begin{tabular}{|c|c||c|c|}
  \hline 
$L$       &    $\nu_B$    &   $\beta_{\nu_B,\infty}$ (small time regime)  & $\beta_{\nu_B,\infty}'$ (intermediate time regime)                      
\\[1mm]
\hline \hline
$3$          &      $1$        &   $\frac{392}{729} \simeq 0.5377$ & 
\\[1mm]
\hline
$4$          &      $1$        &   $ \frac{26333}{49152} \simeq 0.5357$ & 
\\[1mm]
\hline
$5$          &      $1$        &   $\simeq 0.5344$ & 
\\[1mm]
\hline
$\infty$     &     $1$        &    $\simeq 0.5287$ &  $\simeq 0.6032$
\\[1mm]
\hline
$3$          &      $2$        &   $\simeq 0.5741$ & 
\\[1mm]
\hline
$\infty$     &    $2$    &          $\simeq 0.5710$ &  $\simeq 0.6381$
\\[1mm]
\hline
\end{tabular}
  \caption{\label{tab-1} Proportionality factors $\beta_{\nu_B,\infty}$ of Eq. (\ref{eq-proportionality_Schmidt_number_relative_current2}) 
    and $\beta_{\nu_B,\infty}'$ of Eq. (\ref{eq-time_averaged_Schmidt_number_intermediate_times_SF_regime}) 
    when the gas $B$ is in the SF regime
    for some values of the site number $L$ and filling factor
  $\nu_B$.}
\end{table}
\end{center}


\subsection{Gases $A$ and $B$ in the MI and SF regime} \label{eq-entanglement_B_SF_regime}

Let us now assume that the gases $A$ and $B$ are in the MI and SF regimes, respectively ($\lambdaA \ll 1 \ll \lambda_B$).
We will show that, after averaging over time, the { shifted} Schmidt number { and 2-R\'enyi entropy of entanglement show} again a
{  linear growth with the system size $L$} and a proportionality
to the relative (time-averaged) $B$-current reduction, 
\begin{equation} \label{eq-proportionality_Schmidt_number_relative_current2}
\langle S_{AB}^{(2)} \rangle_t/L =   \langle {\Kk}_{A B}\rangle_t/L =  \beta_{\nu_B,\infty}  \langle \Jc_B \rangle_t
 \end{equation}
with $ \beta_{\nu_B,\infty}$ depending on the filling factor $\nu_B$ and being almost independent of $L$.
For instance, for $\nu_B=1$ one has $\beta_{\nu_B,\infty} = 0.53\pm 0.01$ with a difference between the values 
for $L=3,4$, $5$, or $L \to \infty$ smaller than  $0.01$ (see Table~\ref{tab-1}). 
To prove (\ref{eq-proportionality_Schmidt_number_relative_current2}),
we assume that the Peierls phase $\phi_B$ is not close to a half-integer value of $\phi_0$
and approximate the GS $\ket{\psi_B}$ of
the gas $B$ by the SF state
\begin{equation} \label{eq-SF-state}
  \ket{\psi_\SF} = \frac{1}{\sqrt{L^\NB \NB !}} \bigg( \sum_{j=0}^{L-1} e^{\I \phi_0 \ell j} b_j^\dagger \bigg)^\NB \ket{0}
  =  \sqrt{\frac{\NB !}{L^\NB }} \sum_{\nv} \frac{ e^{\I \phi_0 \ell \sum_j j n_j}}{\sqrt{n_0!\cdots n_{L-1}!}} \ket{\nv}
\end{equation}
(here $\ket{0}$ denotes the vacuum state), making a small error $\Oo (\lambdaB^{-1} )$.
As mentioned above, in the short time regime $t \ll J_B^{-1}, U_A^{-1}$ one may replace $\WB$ by
$-V \dnopB$ in the quantum expectation in  (\ref{eq-Schmidt_number_general_expression}), which becomes 
\begin{eqnarray} \label{eq-Q_expectation_B_SF}
  \nonumber \big\langle \E^{-\I t \WB} \big\rangle_{\psi_B}
  & = & 
  \sum_{\nv} \E^{\I t V (n_1-n_0)} \big| \braket{\nv}{\psi_\SF} \big|^2 + \Oo (\lambdaB^{-1}) \\
  \nonumber
  & = & \frac{\NB !}{L^{N_B}} \sum_{\sigma=0}^{N_B} \bigg( \sum_{n_0+n_1=\sigma} \frac{\E^{\I t V n_1} \E^{-\I t V n_0}}{n_0! n_1!} \bigg) 
      {\sum_{\nv'}}^{(\sigma)} \frac{1}{n_2'! \cdots n_{L-1}' !} + \Oo (\lambdaB^{-1})
      \\
      & = & \sum_{\sigma=0}^{N_B} \frac{1}{L^\sigma} \Big( 2 \cos (tV) \Big)^\sigma \frac{\NB !}{\sigma! (\NB - \sigma)!} 
      \bigg( \frac{L-2}{L} \bigg)^{\NB-\sigma} + \Oo (\lambdaB^{-1})\;.
\end{eqnarray}
Here, the sum over $\nv'$ in the second line runs over all $(L-2)$-dimensional vectors $\nv'= (\nv_2' ,\cdots, n_{L-1}')\in \pinteger^{L-2}$ such that
$\sum_{j=2}^{L-1} n_j' = \NB - \sigma$, and in the last line we have used the multinomial formula
$$
\bigg( \sum_{j \in \Jj} x_j \bigg)^N = N! \sum_{\sum_j n_j =N, n_j\geq 0}\; \prod_{j \in \Jj}\frac{x_j^{n_j}}{n_j!}
\qquad\text{if
$\Jj \subset \pinteger$ is a finite subset of indices and $x_j \in \real$.} 
$$
Plugging (\ref{eq-Q_expectation_B_SF}) into (\ref{eq-Schmidt_number_general_expression}), this yields
\begin{equation} \label{eq-Schmidt_number_case(ii)}
 \frac{\Kk_{AB} (t) }{L} =  4 \alpha_A \lambdaA^2 \bigg\{ 1 -  \bigg( 1 - \frac{2}{L} \big( 1 - \cos ( t V) \big) \bigg)^{2\NB}  
 + \Oo (\lambdaA ) + \Oo (\lambdaB^{-1})  \bigg\} \;.
 \end{equation}
Averaging over a time interval $[0, t]$ with $t \gg V^{-1}$
and using Eq. (\ref{eq-relation_current-variation_visib}),
one obtains Eq. (\ref{eq-proportionality_Schmidt_number_relative_current2}) above with 
the proportionality factor 
\begin{equation}  \label{eq-beta_B_SF}
\beta_{\nu_B, \infty} = \frac{2}{3}\bigg\{ 1 - \frac{1}{2\pi} \int_0^{2\pi} \D s \, \bigg( 1 - \frac{2}{L} \big( 1 - \cos s \big) \bigg)^{2\NB} \bigg\}   
 \sim \frac{2}{3}\bigg\{ 1 - \frac{1}{2\pi} \int_0^{2\pi} \D s \,e^{-4\nu_B ( 1 - \cos s )} \bigg\} \;,
\end{equation}
where the last expression corresponds to  the thermodynamical limit $N_B, L \gg 1$ with a fixed $\nu_B =\NB/L$. 
The values of $\beta_{\nu_B, \infty}$ for $\NB=3, 4$, and $5$ and $\nuB=1$ or $2$  are given in Table~\ref{tab-1}.
It is noteworthy that they coincide with the value at the thermodynamical limit up to the second decimal.

  The time evolution of the Schmidt number  determined numerically
  when the gas $B$ is in the SF regime
is shown in Figs.~\ref{fig-new_fig}(a) and (b). One sees that
$\Kk_{AB} (t)$ presents,  in addition to oscillations of period $2\pi/V$, a complex behavior which
is likely to result from the superposition of the different frequencies { associated to the energy parameters $U_A,U_B$, and $J_B$
 entering the Bose-Hubbard Hamiltonians}.
 Nonetheless, the time-averaged Schmidt number $\langle \Kk_{AB}\rangle_t$ has a much simpler evolution and
 basically displays two plateaus. The first plateau at small
time $t \lesssim J_B^{-1}$ agrees well with the predicted value obtained by replacing
$\langle \Jj_B \rangle_t$ by $6 \alpha_A \lambda_A^2$ in 
Eq. (\ref{eq-proportionality_Schmidt_number_relative_current2}) (blue dashed line).
We shall derive  in Sec.~\ref{sec-Schmidt_number_intermediate_times}
an analytical formula giving the value of the second plateau
at times $t \gtrsim J_B^{-1}$ 
in the limit $L \to \infty$ (see Eq.~(\ref{eq-summary_formula_Schmdit_number_B_SF}) below).
The fact that this formula reproduces well the value of the second plateau for
$L=4$ in Fig.~\ref{fig-new_fig}(a) is an indication that
 the time-averaged Schmidt number is not sensitive to { finite size} effects.
In Fig.~\ref{fig-new_fig}(c) one observes that $\langle \Kk_{AB} \rangle/L$
is proportional to $\langle \Jj_B \rangle$ for small values of   $\langle \Jj_B \rangle$, \ie,
 small values of $\lambda_A$, with a proportionality factor in very good agreement with the
predicted value $\beta_{\nu_B,\infty}$  in Eq. (\ref{eq-proportionality_Schmidt_number_relative_current2}).
Interestingly, although $\langle \Kk_{AB} \rangle/L$ is seen
to behave non-linearly  with $\langle \Jj_B \rangle$ at larger $\langle \Jj_B \rangle$'s, the curves obtained for $L=3$, $L=4$, and $L=5$ 
are superposed on  each others, suggesting once again that { finite size} effects are not relevant.

{ 
  Note that our calculations of the $B$-current and Schmidt number in the previous and in this section are
  restricted to  times much smaller than the inverse single species
 energies (small time regime). In the thermodynamical limit, since these energies scale linearly with the atom numbers $N_A$ and $N_B$,
  to insure the validity of our results up to  finite times $t>0$ one must a priori
 re-scale the energy parameters in the Bose-Hubbard Hamiltonians $\HA$ and $\HB$  as
 $U_A \to U_A/N_A$, $J_A \to J_A/N_A$, $U_B \to U_B/N_B$ and $J_B \to J_B/N_B$.

 We will extend in Sec.\ref{sec-Schmidt_number_intermediate_times}
 the above results on the Schmidt number  in both the MI regime  $\lambdaB \ll 1$ and the SF regime
 $\lambdaB \gg 1$ to larger times of order $U_B^{-1}$ and $J_B^{-1}$, respectively, keeping the assumption $t$ much smaller than the
 inverse interaction energy of the $A$-atoms.
 It will be shown that the time-averaged Schmidt number is still given at these larger times
 by Eqs.~(\ref{eq-average_Schmidt_number_small_lambdaA_lambdaB})
 in the MI regime and by Eq.~(\ref{eq-proportionality_Schmidt_number_relative_current2})
 with a sightly different prefactor $\beta_{\nuB,\infty}$ in the SF regime.}
 
\begin{figure}
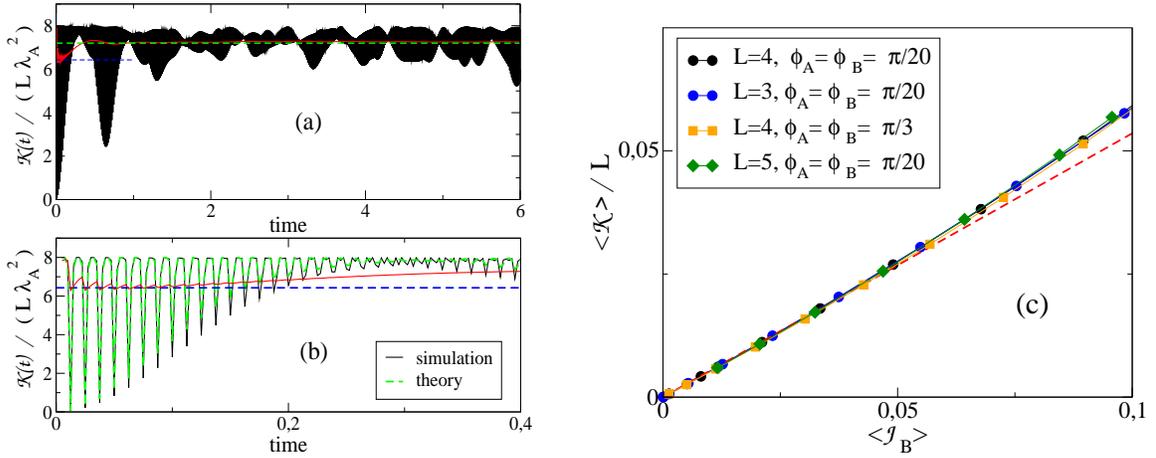

    \begin{center}
      \includegraphics[width=7cm, height=6cm, angle = 0]{Fig5a.eps}
      \hspace{0.5cm}
      \includegraphics[width=7.4cm, angle = 0]{Fig5b.eps}
    \end{center}
    \caption{{\bf (a):}
      Schmidt number $\Kk_{AB} (t)$
      divided by $L \lambda_A^2$  vs.
      time  (black plain line)
      and its time average  $\langle \Kk_{AB} \rangle_t/(L \lambdaA^2)$ (red plain line) for a gas $B$ in the SF regime.
      Time is in units of $U_B^{-1}$.
      The parameters are $L=N_A=N_B=4$, $V=500 U_B$, $J_B=5 U_B$, $U_A=0.1 U_B$, $J_A=0.002 U_B$, and
      $\phi_A=\phi_B=\pi/3$. The blue and green dashed horizontal lines display
      the time averaged  values from Eq.~(\ref{eq-summary_formula_Schmdit_number_B_SF}) in the short and intermediate time regimes, respectively.
      {\bf (b):} same on the time interval $[0,0.4/U_B]$ with
      formula (\ref{eq-Schmidt_number_intermediate_times_SF_regime}) shown in green dashed lines.
      {\bf (c):} Time-averaged Schmidt number per site $\langle \Kk_{AB} \rangle_t/L$ vs.
      relative current reduction $\langle {\cal J}_B \rangle_t$, for $N_A=N_B=L$ and  $\phi_A=\phi_B=\phi$
      with      $L= 3$, $\phi=\pi/20$ (blue dots), $L=4$,  $\phi =\pi/20$ (black dots),
      $L=4$, $\phi=\pi/3$ (orange squares), and $L=5$, $\phi=\pi/20$ (green diamonds). The energy parameters are  
      $U_A=0.1 U_B$, $J_B=U_B$, $J_A$ variable,  averaging time $t=0.3/U_B$.  
      Dashed line: values from Eqs. (\ref{eq-proportionality_Schmidt_number_relative_current2}) with $\beta_{\nu_B,\infty}=0.54$ (see Table~\ref{tab-1}).
     }
  \label{fig-new_fig}
\end{figure}

\subsection{Gas $A$ in the MI regime, $\lambda_B$ arbitrary} \label{entanglement_arbitary_lambda_B}

In the more general case $\lambda_B$ arbitrary and $\lambda_A \ll 1$, one deduces from (\ref{eq-Schmidt_number_general_expression}) and
(\ref{eq-relation_current-variation_visib}) that
\begin{equation} \label{eq-proportionality_Schmidt_number_relative_current3}
  \langle {\Kk}_{A B}\rangle_t/L =  \beta_{\nu_B,\lambda_B,L}  \langle \Jc_B \rangle_t
  \quad , \quad V^{-1} \ll t \ll U_A^{-1}, U_B^{-1}, J_B^{-1}\;,
 \end{equation}
with a proportionality factor given by 
\begin{equation}
  \beta_{\nu_B,\lambda_B,L}
  = \frac{2}{3}
  \bigg( 1 - \frac{1}{2\pi} \int_0^{2\pi} \D s  \big| \big\langle \E^{\I s \dnopB} \big\rangle_{\psi_B} \big|^2 \bigg)
  = \frac{2}{3}
  \bigg( 1 -
  \sum_{\nv,\mv,n_1-n_0=m_1-m_0} | \braket{\nv}{\psi_B} |^2 | \braket{\mv}{\psi_B} |^2 
  \bigg)
  \;.
\end{equation}
Note that $\ket{\psi_B}$ and thus $ \beta_{\nu_B,\lambda_B,L}$  depend on $L$, so that we cannot claim that
Eq. (\ref{eq-proportionality_Schmidt_number_relative_current3}) shows a {  linear growth of entanglement  with the system size} for finite lattices
and $\lambda_B \approx 1$, although one may expect that $ \beta_{\nu_B,\lambda_B,L}$ converges rapidly to its large $L$ limiting value
as in the case $\lambda_B \gg 1$.

Since only observables and quantum expectations of the gas $B$ appear in
the RHS of Eq.~(\ref{eq-proportionality_Schmidt_number_relative_current3}),
this equation provides a way to estimate the amount of entanglement
between the two gases when $\lambdaA \ll 1$ by {\it performing local measurements on the species $B$.}

\section{Quantum superpositions at the origin of the entanglement} \label{sec-superpositions}

{

 In order to understand better the entanglement generation process in the quench dynamics, it is enlightening to
  look at the superpositions at the origin of this entanglement in the time-evolved wavefunction of the two gases. 
  With this aim, we determine this wavefunction $\ket{\psi_{AB} (t)}$ to second order in $\lambda_A$.

  Disregarding as before the
  tunneling Hamiltonian $\KA$ and the commutator $[ \hat{H}_{AB}^{\rm int} , \HB]$ in the dynamical evolution up to times $t \ll U_A^{-1}, U_B^{-1}, J_B^{-1}$
  and recalling that $\HB \ket{\psi_B} = E^B_{\rm GS} \ket{\psi_B}$,
  we find from (\ref{eq-HAB_int}) and  (\ref{eq-pertubrative_expansion_GS})
  \begin{equation}
   \ket{\psi_{AB} (t)} = \E^{-\I t \hat{H}_{AB}^{\rm int}} \, \E^{-\I t \hat{H}_A^{\rm int}} \, \E^{-\I t E_{\rm GS}^B}
   \big( \ket{\psi_\MI}+ \lambdaA \ket{\psi_A^{(1)}} + \lambdaA^2 \ket{\psi_A^{(2)}} + \Oo (\lambdaA^3)  \big) \ket{\psi_B}\;.
  \end{equation}
Let us expand the initial GS of $B$ in the Fock basis as
$\ket{\psi_B}= \sum_{\nv_B} c_{\nv_B} \ket{\nvB}$
and separate the second-order correction to the GS of $A$ as the sum of 
$\braket{\psi_\MI}{\psi_A^{(2)}} \ket{\psi_\MI}$ and $\Pi_\MI^\bot \ket{\psi_A^{(2)}} $, where $\Pi_\MI^\bot = 1 - \ketbra{\psi_\MI}{\psi_\MI}$ is the projector
onto the orthogonal subspace to the MI state. Using  (\ref{eq-1st_order_corr_GS}) and (\ref{eq-2nd_order_correc_GS}) and recalling
that the energy to create a particle-hole excitation is equal to $U_A$, we obtain
  \begin{eqnarray} \label{eq-time_evaloved_state_AB}
    \nonumber
   \ket{\psi_{AB} (t)}  & = & 
    \E^{-\I t ( V \nu_A N_B + E^{(0)}_A (\nu_A) + E_{\rm GS}^B)} \bigg\{  ( 1 -\lambdaA^2 \alpha_A L ) \ket{\psi_\MI} \ket{\psi_B} 
      \\ \nonumber
    &   & 
   + \lambda_A \sqrt{\alpha_A} \, \E^{-\I t U_A} \sum_j \sum_{\nv_B} c_{\nv_B}   \Big(
    \E^{\I \phi_A} \E^{-\I t V ( n_{j+1}^B- n_j^B )} \ket{\varphi_{A,j,+}}  +  \E^{-\I \phi_A} \E^{\I t V ( n_{j+1}^B- n_{j}^B )} \ket{\varphi_{A,j,-}} \Big) \ket{\nv_B} \bigg\}
    \\
    &   &
    + \lambdaA^2 \E^{-\I t E_{\rm GS}^B}  \,\E^{-\I t \hat{H}_{AB}^{\rm int}} \, \E^{-\I t \hat{H}_A^{\rm int}} \, \Pi_\MI^\bot  \ket{\psi_A^{(2)}}  \ket{\psi_B}
    + \Oo (\lambda_A^3 ) \; ,
\end{eqnarray}
  where $\ket{\varphi_{A,j,\pm}}$ are the particle-hole states, see~\ref{eq-particle-hole_state}.
  It is easy to convince oneself  that the contribution of the term in the last line of (\ref{eq-time_evaloved_state_AB}) to  the reduced density matrix
  $\hat{\rho}_B (t)=\tr_A \ketbra{\psi_{AB}(t)}{\psi_{AB}(t)}$
  of the gas $B$ is of order $\lambdaA^3$ or higher.
  Since we are interested in the Schmidt number and the $B$-current, which both depend on $\hat{\rho}_B (t)$ only
  (see (\ref{eq-def_Schmidt_number}) and (\ref{eq-general_expression_B_current})), this term can be
  dropped out. 
 Disregarding also the irrelevant phase factor, the wavefunction of the two gases at time $t$ is given by
  \begin{equation}  \label{eq-time_evaloved_state_ABbis}
    ( 1 -\lambdaA^2 \alpha_A L )  \ket{\psi_\MI} \ket{\psi_B} + \lambda_A \sqrt{\alpha_A}\, \E^{-\I t U_A} \sum_{j, \pm}
    \E^{\pm \I \phi_A} \ket{\varphi_{A,j,\pm}}  \ket{\psi_{B,j,\pm}(t)} + \Oo (\lambdaA^3)
  \end{equation}
  with
  \begin{equation} \label{eq-psi_B_j_pm}
    \ket{\psi_{B,j,\pm} (t)}  =  \sum_{\nvB} c_{\nvB} \E^{\mp \I t V ( n_{j+1}^B - n_j^B)} \ket{\nv_B}\;.
  \end{equation}
  The corrections of order $\lambda_A$  to the separable state $ \ket{\psi_\MI} \ket{\psi_B}$ in (\ref{eq-time_evaloved_state_ABbis})
  are at the origin of the entanglement between the two gases. Thus entanglement comes from the coupling of particle-hole excitations
  $\ket{\varphi_{A,j,\pm}}$ in the gas $A$ with the states
  $\ket{\psi_{B,j,\pm} (t)}$ of the gas $B$ at each lattice site $j$.
  Let us note that the latter states  are periodic in time with period $2\pi/V$; in particular, they come back
to  their initial $j$-independent value $\ket{\psi_B}$ at times $t_m = 2 m \pi/V$, $m=1,2,\ldots$,
  leading to a disappearance of entanglement (this is valid when $t_m \ll U_A^{-1},U_B^{-1},J_B^{-1}$; see the next section for larger times).
  This explains the observed oscillations
  of $\Kk_{AB} (t)$ with period $2\pi/V$ in the inset of Fig.~\ref{Fig-3}(a) and in Fig.~\ref{fig-new_fig}(b).

  We now look at the effect of the aforementioned superpositions on the $B$-current $I_B (t) = \tr_B [ \hat{\rho}_B (t) \IB ] $.
  Plugging into this formula the reduced density matrix
  $\hat{\rho}_B(t)$ obtained from the wavefunction~\ref{eq-time_evaloved_state_ABbis}, one gets
  \begin{equation} \label{eq-B-current_new_formula}
    I_B (t) =   ( 1 -2 \lambdaA^2 \alpha_A L ) I_B (0) +  \lambdaA^2 \alpha_A \sum_{j,\pm} \big\langle \hat{I}_B \big\rangle_{\psi_{B,j,\pm} (t)} + \Oo (\lambdaA^3)\;.
  \end{equation}
  The reduction of the $B$-current comes from the fact that the currents $\langle \hat{I}_B \rangle_{\psi_{B,j,\pm} (t)}$ in the
  states $\ket{\psi_{B,j,\pm} (t)}$ are smaller than $I_B (0)$.
In fact, a simple calculation shows that these currents averaged in the time interval $[0,t]$ with $t \gg 1/V$ are
equal to
\begin{equation}  \label{eq-B-average_current_new_formula}
  \big\langle \langle \hat{I}_B \big\rangle_{\psi_{B,j,\pm}} \big\rangle_t
  = \frac{1}{2 \I L} \sum_{i\notin \{ j-1, j,j+1\} } \bra{\psi_B} \big( \E^{\I \phi_B} \hat{b}_{i+1}^\dagger \hat{b}_i - {\rm h.c.} \big) \ket{\psi_B} \;.
\end{equation}
This  expression is nothing but the current for the initial state $\ket{\psi_B}$ in the ring after having removed the three sites $i=j- 1$, $j$, and  $j+1$.
Replacing $\langle \hat{I}_B \big\rangle_{\psi_{B,j,\pm} (t)}$ by $(L-3) I_B(0)/L$ in
(\ref{eq-B-current_new_formula}), one gets back formula (\ref{eq-relation_current-variation_visib}) for the relative current variation.
We conclude that the universal reduction of $B$-current is due to the coupling of particle-hole excitations in the gas $A$ with the
states (\ref{eq-psi_B_j_pm}) of the $B$-gas. The latter have a time-averaged current reduced by a factor $(L-3)/L$, in qualitative
agreement with the intuitive energetic argument given in Sec.~\ref{sec-supercurrent}.

 Let us stress that as a consequence of the energy scale separation (\ref{eq-separation_of_timescales}),
  all states appearing in the wavefunction (\ref{eq-time_evaloved_state_ABbis}) have energies much higher than the GS energy  $E_{AB}^{\rm GS} \approx-V N_A N_B$ for the post-quench Hamiltonian $\hat{H}_{AB}$.
  More precisely,
  \begin{eqnarray} \label{eq-energies_states_in_superposition}
    \big\langle \hat{H}_{AB} \big\rangle_{\psi_\MI \otimes \psi_B}
    & \equiv  & - V \nu_A N_B +  E_A^{(0)} (\nuv_A) + E_{\rm GS}^B + \Oo ( J_A) \\ \nonumber
    \big\langle \hat{H}_{AB}\big\rangle_{ \varphi_{A,j,\pm}  \otimes \psi_{B,j,\pm} (t)}
    & = & - V \nu_A N_B  - V  \big\langle \hat{n}_{j+1}^B- \hat{n}_j^B  \big\rangle_{\psi_B}
    + E_A^{(0)} (\nuv_A) + U_A  +  \big\langle \HB \big\rangle_{\psi_{B,j,\pm} (t)} + \Oo ( J_A)\;.
  \end{eqnarray}
In particular, the GS of the post-quench Hamiltonian $\hat{H}_{AB}$ does not play any role in the quench dynamics at the times we are considering.
  This can be explained from the fact that the initial state $\ket{\psi_A} \ket{\psi_B}$ has an energy
  lying in the highly excited part of the spectrum of  $\hat{H}_{AB}$.
  This is illustrated in the numerical results of  Fig.~\ref{Fig_9} displaying the probabilities
  $p_i = | \braket{\psi_{AB}(t)}{\Phi_i}|^2$ of finding the wavefunction of the two gases
  in the eigenstates $\ket{\Phi_i}$ of $\hat{H}_{AB}$  (note that the $p_i$ are time independent).
  The peak with value $p_i^\ast$ close to 1 corresponds to the eigenstate of $\hat{H}_{AB}$   with closest eigenvalue to
$E_{AB}^{(0) \;\ast} =  - V \nu_A N_B +  E_A^{(0)} (\nuv_A) + E_{\rm GS}^B$.
Actually, the  state $\ket{\psi_\MI} \ket{ \psi_B}$  in (\ref{eq-time_evaloved_state_ABbis})
is an eigenstate of   $\hat{H}_{AB}^{(0)} = \HA^{\rm int} + \HB + \hat{H}_{AB}^{\rm int}$
  with eigenvalue $E_{AB}^{(0) \; \ast}$. Thus by perturbation theory $\hat{H}_{AB}=\hat{H}_{AB}^{(0)} + \KA$ has an eigenvector
  $\ket{\Phi_{AB}^\ast}$ differing from this state by $\Oo ( J_A/V)$ with  eigenenergy
  $E_{AB}^\ast = E_{AB}^{(0) \ast} + \Oo ( J_A)$.
  The corresponding  probability $p_i^\ast$ is the square of the coefficient multiplying $\ket{\psi_\MI} \ket{\psi_B}$ in (\ref{eq-time_evaloved_state_ABbis}), 
  \begin{equation}
    p_i^\ast = \big| \bra{\psi_\MI} \braket{\psi_B}{\psi_{AB} (t)} + \Oo ( J_A/V  ) \big|^2 
     =  1 -2 \lambdaA^2 \alpha_A L + \Oo (\lambdaA^3) + \Oo (J_A/V)\; .
  \end{equation}
  The small non-zero probabilities shown in the two insets
  are the contribution of the linear terms in (\ref{eq-time_evaloved_state_ABbis}), which are at the origin of the entanglement.
  The peak in the upper inset comes from the time-independent contribution in these linear terms
  obtained by forgetting all Fock states with $n_j^B \not= n_{j+1}^B$
  in the definition (\ref{eq-psi_B_j_pm}) of $\ket{\psi_{B,j,\pm} (t)}$; such states
  have inter-species interaction energy $-V \nu_A N_B$ and
  contribute to the time-averaged current.     
}

\begin{figure}
    \begin{center}
      \includegraphics[width=7cm, angle=0]{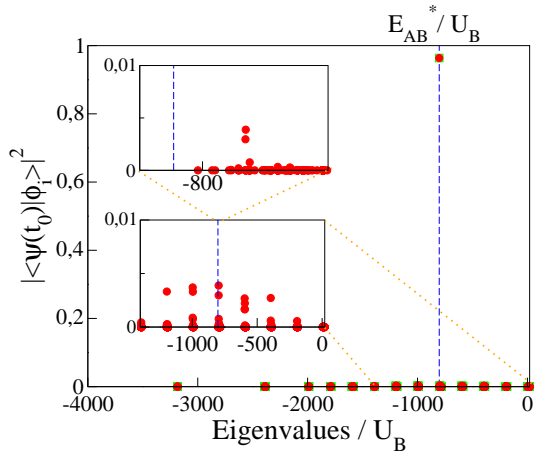}
    \end{center}
    \caption{
      Probabilities $p_i = | \braket{\psi_{AB}(t_0)}{\Phi_i}|^2$ of finding the wavefunction of the two gases
      at time $t_0$ in an eigenstate $\ket{\Phi_i}$ of the post-quench Hamiltonian $\hat{H}_{AB}$, for
      $N_A=N_B=L=4$, $V=200 U_B$, $J_B=U_A=U_B$, $J_A=0.05 U_B$, $\phi_A=\phi_B=\pi/10$, and $t_0 = 1/U_B$  (from numerical exact diagonalization). We checked that the numerically evaluated values of
 $p_i$ do not change when changing $t_0$, as it should be.
      Each red point corresponds to a 
      given eigenstate of  $\hat{H}_{AB}$ with an eigenenergy $E_{AB}^i$ represented in the horizontal axis (in units of $U_B$)
      and the associated probability $p_i$ represented in the vertical axis. 
      The eigenenergy $E_{AB}^\ast$ at which $p_i$ displays a maximum
      is defined in the text.  The insets show amplifications in particular regions of the spectrum. For energies
      $E_{AB}^i \lesssim -1600 U_B$ the values of $p_i$ 
      are smaller than $2 \times 10^{-5}$.     
    }
    \label{Fig_9}
\end{figure}

 \section{Interspecies entanglement at larger times}
\label{sec-Schmidt_number_intermediate_times}

 In this  section, we determine the Schmidt number at times $t$ that can be of order
$U_B^{-1}$ or $J_B^{-1}$ but must
be small with respect to the inverse interaction energy of the gas $A$, \ie,
\begin{equation} \label{eq-intermediate_times}
0 \leq   t \ll  (N_A U_A)^{-1} \;.
\end{equation}  
Having in mind that the interaction energy  of the gas $A$ can be tuned to be smaller than $U_B$ and $J_B$
 (this can be done experimentally using Feshbach resonances or by varying the atom number $N_A$),
we define an {\it intermediate time regime} by $U_{B}^{-1},  J_B^{-1} \lesssim t \ll ( N_A U_A)^{-1}$.
Note that one may exchange the role of $A$ and $B$ in the
previous inequalities, using  the symmetry property of $\Kk_{AB}(t)$.
The main results established below are:
\begin{itemize}
\item[(i)]  Formula  (\ref{eq-time_evolution_of_K}) generalizes Eq. (\ref{eq-Schmidt_number_small_lambdaA_lambdaB}) to larger times,
  giving the time evolution of $\Kk_{AB}(t)$ when  the gas $B$ is in the MI regime.
\item[(ii)] Under the same condition, the time-averaged shifted Schmidt number $\langle \Kk_{AB} \rangle_t$ 
  is still given by
  Eq.~(\ref{eq-average_Schmidt_number_small_lambdaA_lambdaB}) in the intermediate time regime. In the numerical simulations shown in Fig.~\ref{Fig_app3}(a),
  one indeed observes that $\langle \Kk_{AB} \rangle_t$ stays constant over a wide range of time, even at 
  times larger than $U_A^{-1}$.
\item[(iii)]  Formula (\ref{eq-Schmidt_number_intermediate_times_SF_regime}) generalizes 
  Eq. (\ref{eq-Schmidt_number_case(ii)}) to  larger times,
  giving the time evolution of $\Kk_{AB}(t)$ when the gas $B$ is in the SF regime.
\item[(iv)] Under the same condition, Eq. (\ref{eq-summary_formula_Schmdit_number_B_SF})
  shows that the time-averaged shifted Schmidt number $\langle \Kk_{AB} \rangle_t$ takes
  in the intermediate time regime  a constant value slightly above the short-time value; we find numerically that
  $\langle \Kk_{AB} \rangle_t$ is constant even at times larger than $U_A^{-1}$ and $U_B^{-1}$, see Fig.~\ref{Fig_app3}(b). 
\end{itemize}
The analytical calculations leading to these results are technically more involved than those of the  previous sections.
Actually, to be able to estimate  $\Kk_{AB} (t) $ in the intermediate time regime one
needs to determine the spectrum and eigenvectors of the Hamiltonian $\WB$
appearing in Eq.~(\ref{eq-Schmidt_number_general_expression}). This is done
in the two
limiting cases $\lambda_B \ll 1$ (gas $B$ in the MI regime) and
$\lambda_B \gg 1$ (gas $B$  in the SF regime) in the two next subsections.

\begin{figure}
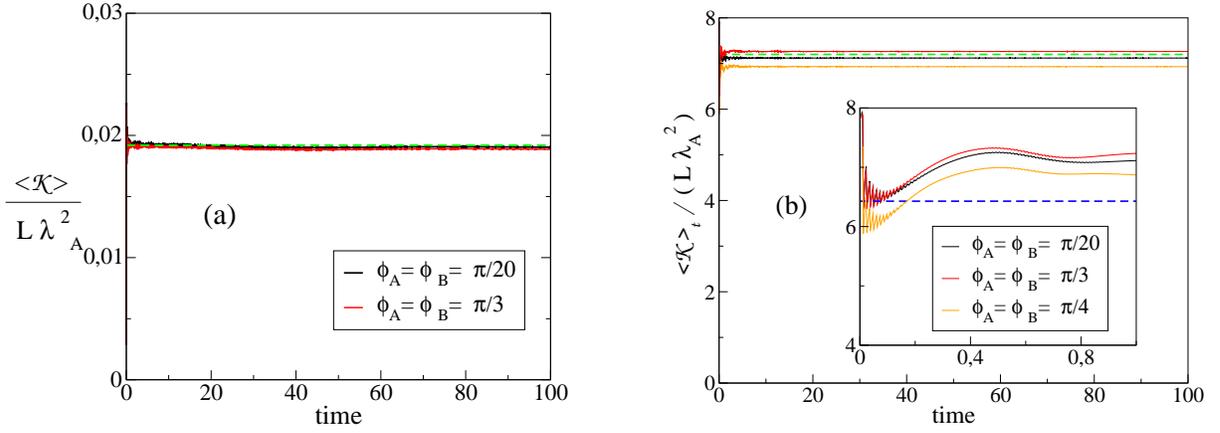

    \begin{center}
      \includegraphics[width=7.5cm, angle=0]{Fig6a.eps}
      \hspace{1cm}
      \includegraphics[width=7.2cm, angle =0]{Fig6b.eps}
    \end{center}
    \caption{{\bf (a):}
      Time-averaged Schmidt number per site $\langle{\Kk_{AB}} \rangle_t/L$ divided by $\lambdaA^2$
      as function of time $t$ (in units of $U_B^{-1}$) for a gas $B$ in the MI regime up to $t=100/U_B$
      (from numerical calculations).
      Parameters: $L=N_A=N_B=4$, $V=100 U_B$, $J_B=0.01 U_B$, $U_A=0.1 U_B$, and $J_A=0.001 U_B$, $\phi_A=\phi_B=\pi/20$
      (black line) and $\pi/3$ (red line). Horizontal green dashed line:
      value from Eq.~(\ref{eq-average_Schmidt_number_small_lambdaA_lambdaB}). 
      {\bf (b):} Same  for a gas $B$ in the SF regime.  
      Parameters: $L=N_A=N_B=4$, $V=500 U_B$, $J_B=5 U_B$, $U_A=0.1 U_B$, $J_A=0.001 U_B$,
      $\phi_A=\phi_B=\pi/20$ (black line), $\pi/3$ (red line) and $\pi/4$ (orange line), with the
      behaviors at times between $0$ and  $U_B^{-1}$ shown in the inset.
      The horizontal blue and green dashed lines display the values from
      Eq.~(\ref{eq-summary_formula_Schmdit_number_B_SF}) in the short and  intermediate time regimes, respectively.
      Note that the phase $\pi/4$ corresponds to the critical value $\phi_0/2$ for which
      Eq.~(\ref{eq-summary_formula_Schmdit_number_B_SF}) is not valid (see text).
    }
    \label{Fig_app3}
\end{figure}

\subsection{Gases $A$ and $B$ in the MI regime} \label{sec-Schmidt_number_case(i)}

As before, we consider a gas $A$ initially in the MI regime
and inter-species interaction strengths $V$
much larger than the single species energies, see~(\ref{eq-separation_of_timescales}).
We assume in this subsection that, in addition, the gas $B$ is also in the MI regime, i.e. $\lambdaB \ll 1$.
We determine the shifted Schmidt number in these limits by relying on
formula (\ref{eq-Schmidt_number_general_expression}), which gives a good approximation of $\Kk_{AB} (t) $
at times $t$ satisfying (\ref{eq-intermediate_times}), as shown in Appendix~\ref{app-B}.
Using the spectral decomposition of $\E^{-\I t \Wop_B}$, this gives
\begin{equation} \label{eq-kk_after_spectral_decomp}
  \Kk_{AB} (t) = 4 L \alphaA \lambdaA^2 \bigg[ 1 - \sum_{\nv,\nv'} \cos \big( t (w_\nv-w_{\nv'} ) \big)
    | \braket{\phi_\nv}{\psi_B } |^2  | \braket{\phi_{\nv'}}{\psi_B } |^2 \bigg]\;,
\end{equation}
where $\ket{\phi_\nv}$ and $w_\nv$ are the eigenstates and eigenenergies of the Hamiltonian
$\WB$.
In order to estimate these eigenstates and energies, we diagonalize $\WB$ through a two-fold application of
time-independent perturbation theory.

Firstly, we treat the Bose-Hubbard Hamiltonian $\HB$ of the gas $B$ as a perturbation of $-V \dnopB$. The latter (unperturbed)
Hamiltonian has degenerated eigenvalues $w_\Delta^{(0)} = - V \Delta$ and  eigenprojectors
\begin{equation} \label{eq-projector_Pi_Delta}
  \PiB = \sum_{\nv, (\Delta \nv)_{01} = \Delta} \ketbra{\nv}{\nv}
\end{equation}
with $\Delta \in \{ - N_B,\ldots, N_B\}$, where we have set $(\Delta \nv)_{01}\equiv n_1-n_0$.
Hereafter, we omit the label $B$ referring to the gas $B$ to simplify notations.
To lowest order in $U_B/V$, the eigenstates of $\Wop$ are given by the eigenstates of the projected Bose-Hubbard Hamiltonian
\begin{equation} \label{eq-projected_Bose_Hubbard}
  \PiB \Hop \,\PiB = \Hop^\inter \PiB + \PiB \Kop \PiB
\end{equation}
and the corresponding eigenvalues are the  first-order corrections  to the unperturbed energies $w_\Delta^{(0)}$.
Recall that the kinetic Hamiltonian reads
\begin{equation} \label{eq-kinetic_part_BH}
  \Kop =  - J_B\sum_{<i,j>}  \E^{\I \phi_{ij}} \bop^\dagger_{j} \bop_i  \;,
\end{equation}
where  $<i,j>$ refers to nearest neighbor sites $i,j$ on the ring lattice and $\phi_{ij} = \pm \phi_B$ if
$j=i\pm 1$.  

For $\lambdaB \ll 1$,  one may diagonalize the Hamiltonian  (\ref{eq-projected_Bose_Hubbard}) using once again perturbation theory,
now with the intraspecies interaction  term $\Hop^\inter \PiB$ as the unperturbed part and the projected kinetic Hamiltonian
$\PiB \Kop \PiB$ as the perturbation. Denote as before  by $E^{(0)}(\nv)$ and by $E^{(0)}_0$ the eigenenergies
 and GS energy of $\Hop^\inter$.
For any  $\Delta \in \{ -N_B,\ldots, N_B\}$ and  any energy
$\epsilon$ in the spectrum of $\Hop^\inter \PiB$, let us introduce the set
$\Ss_{\Delta,\epsilon} = \{ \nv \in \pinteger^{L} \,;\, (\Delta \nv)_{01} = \Delta , \Eint ( \nv ) = \epsilon , \sum_j n_j=N_B\}$ and
let $(c_{\mv,\nv})_{\mv\in \Ss_{\Delta, \epsilon}}$ be the normalized eigenvectors of the matrix $( \bra{\pv} \Kop \ket{\mv})_{\pv,\mv \in \Ss_{\Delta,\epsilon}}$ with eigenvalues
$E^{(1)} ( \nv)$, $\nv \in \Ss_{\Delta,\epsilon}$. The eigenenergies and eigenstates of $\Wop$ are given by
\begin{eqnarray} \label{eq-perturbed_eigenenergies_W}
  w_\nv & = & - V (\Delta \nv)_{01} + \Eint (\nv)  + E^{(1)} (\nv) + \Oo ( \lambdaB  J_B ) + \Oo \Big( \frac{U_B^2}{V} \Big)
  \\
  \label{eq-perturbed_eigensates_W}
  \ket{\phi_\nv} & = & \ket{\phi_\nv^{(0)}} + \lambdaB \ket{\phi_\nv^{(1)}} + \lambdaB^2 \ket{\phi_\nv^{(2)}} + \Oo (\lambdaB^3) + \Oo \Big( \frac{U_B}{V} \Big)
\end{eqnarray}
with $( \Delta \nv)_{01} =\Delta$ and $E^{(0)}(\nv)=\epsilon$, where 
\begin{equation} \label{eq-eigenstates_WB_pertub_theory}
  \ket{\phi_\nv^{(0)}}  =  \sum_{\mv \in \Ss_{\Delta,\epsilon}} c_{\mv,\nv} \ket{\mv} 
\; , \qquad
\lambdaB  \ket{\phi_\nv^{(1)}}  =  \sum_{\pv, (\Delta \pv)_{01} = \Delta, \Eint ( \pv) \not= \epsilon}
 \frac{\ketbra{\pv}{\pv} \Kop \ket{\phi_\nv^{(0)}}}{\Eint (\nv)  - \Eint (\pv)}\;.
\end{equation}  
In particular, since the GS of  $\Hop^\inter$ is non-degenerated and given by
the MI state $\ket{\psi_\MI}=\ket{\nuv}$, the set $\Ss_{\Delta,\epsilon}$ reduces to a single element $\nuv$ 
when $\Delta =0$ and $\epsilon = \Eint_0$. Then (compare with (\ref{eq-1st_order_corr_GS}))
\begin{equation} 
  \ket{\phi_\nuv^{(0)}}  =  \ket{\psi_\MI}
  \; , \qquad 
  \ket{\phi_\nuv^{(1)}}   
      = \sqrt{\alphaB} \sum_{<i,j>, \Delta_{01}^{ij}=0} \E^{\I \phi_{ij}^B} \ket{\nuv + 1_j - 1_i}\;,
\end{equation}
where we have set 
\begin{equation} \label{eq-values_Delta_1_j-1_i}
  \Delta_{01}^{ij} =  - \Delta_{01}^{ji} = \delta_{1,j}-\delta_{1,i}-\delta_{0,j}+\delta_{0,i}  =
  \begin{cases}
    2 & \text{ if $i=0$, $j=1$} \\
    1 & \text{ if  ($i=0$, $j \notin \{0,1\}$) or ($i \notin \{0,1\}$, $j=1$)}\\
      0 & \text{ if $i,j \notin \{ 0,1\}$.}
  \end{cases}
\end{equation}
Furthermore, the second-order correction is given when $\Delta =0$ and $\epsilon = \Eint_0$ by
\begin{eqnarray} \label{eq-2d_order_correction_phi_nu}
  \nonumber
  \lambda^2_B \ket{\phi_\nuv^{(2)}} & = &
  \sum_{\mv, \pv \not= \nuv, (\Delta \mv)_{01}=(\Delta \pv)_{01}=0}
  \frac{\bra{\mv} \Kop \ket{\pv} \bra{\pv} \Kop \ket{\psi_\MI}}{( E_0^{(0)} - E^{(0)} ( \mv)) ( E_0^{(0)} - E^{(0)} ( \pv))} \ket{\mv}
  \\
  & & - \onehalf \sum_{\mv \not= \nuv, (\Delta \mv)_{01}=0} \frac{ | \bra{\mv} \Kop \ket{\psi_\MI} |^2}{( E_0^{(0)} - E^{(0)} ( \mv))^2} \ket{\psi_\MI}\;.
\end{eqnarray}

The next step is to calculate the scalar products $\braket{\phi_{\nv}}{\psi}$ appearing in
(\ref{eq-kk_after_spectral_decomp}) up to second order in $\lambda_B$. To this end, we combine (\ref{eq-perturbed_eigensates_W})
with the perturbative expansion of the GS $\ket{\psi}$ of the gas $B$ analog to (\ref{eq-pertubrative_expansion_GS}).
Since $\Kop$ only couples the MI state to Fock states with a single particle-hole excitation and the energy to create such an excitation
is equal to $U_B$, one has $\bra{\mv} \Kop \ket{\psi_\MI} \not= 0$ if and only if $E^{(0)} ( \mv) = E^{(0)}_0 + U_B$.
Consequently, by (\ref{eq-eigenstates_WB_pertub_theory}),
\begin{eqnarray} \label{eq-almost_finished!}
  \nonumber
  \lambda_B \braket{\phi_\nv^{(1)}}{\psi_\MI}
  & = &
  \begin{cases}
    \dss
    \frac{1}{U_B} \bra{\phi_\nv^{(0)}} \Kop \ket{\psi_\MI} & \text{ if $(\Delta \nv)_{01}=0$ and $E^{(0)} ( \nv) = E^{(0)}_0 + U_B$}
    \\[2mm]
    0     & \text{ otherwise}
  \end{cases}
  \\
  & = & 
\begin{cases}
    \dss - \lambda_B \braket{\phi_\nv^{(0)}}{\psi^{(1)}} & \text{ if $(\Delta \nv)_{01}=0$}
    \\[2mm]
    0     & \text{ if $(\Delta \nv)_{01}\not=0$,}
\end{cases}
\end{eqnarray}
where we have used (\ref{eq-1st_order_corr_GS}) (with $A$ replaced by $B$) in the second equality.
Expanding $\ket{\psi}$ and $\ket{\phi_\nv^{(0)}}$ in powers of $\lambdaB$ and using
$\braket{\phi_\nv^{(0)}}{\psi_\MI} = \delta_{\nv, \nuv}$, $\braket{\phi_\nuv^{(1)}}{\psi_\MI}=\braket{\psi^{(1)}}{\psi_\MI}=0$, and (\ref{eq-almost_finished!}), we obtain 
\begin{equation} \label{eq-scalar_prod_1}
  \big| \braket{\phi_\nv}{\psi} \big|^2 =\!
  \begin{cases}
    1 + 2 \lambdaB^2 \re \big\{ \braket{\phi_\nuv^{(2)}}{\psi_\MI} \!+ \!\braket{\phi_\nuv^{(1)}}{\psi^{(1)}} \!+\! \braket{\psi_\MI}{\psi^{(2)}} \big\}
    \!+ \! \Oo (\lambdaB^3)\! + \!\Oo (\frac{\lambdaB  U_B}{V} )
    & \!\! \text{if $\nv = \nuv$} \\[2mm]
    \lambdaB^2 | \braket{\phi_\nv^{(0)}}{\psi^{(1)}} |^2 + \Oo (\lambdaB^3 ) + \Oo ( \frac{\lambdaB^2  U_B}{V} )
    & \!\! \text{if $\nv \not= \nuv$, $(\Delta \nv)_{01} \not= 0$} \\[2mm]
    \Oo ( \lambdaB^4 ) + \Oo ( \frac{ \lambdaB^3 U_B}{V} ) + \Oo ( \frac{ \lambdaB^2 U_B^2}{V^2} )
    & \!\! \text{if $\nv \not= \nuv$, $(\Delta \nv)_{01} = 0$.}
  \end{cases}
\end{equation}

We proceed to evaluate the scalar products appearing in the RHS of (\ref{eq-scalar_prod_1}).
One easily shows using (\ref{eq-2d_order_correction_phi_nu}) that, in analogy with (\ref{eq-2nd_order_correc_GS}),
\begin{equation}
  \braket{\psi_\MI}{\phi_{\nuv}^{(2)}} = -\onehalf \sum_{<i,j>, \Delta_{01}^{ji}=0} \alphaB =  - \alphaB ( L  - 3 )\;,
\end{equation}
where we have taken advantage of (\ref{eq-values_Delta_1_j-1_i}) in the last equality.
Furthermore, $\braket{\phi_{\nuv}^{(1)}}{\psi^{(1)}}$ is given by the same expression multiplied by $-2$.  
Replacing these values into (\ref{eq-scalar_prod_1}), this yields
\begin{equation} \label{eq-scalar_prod_2}
 \big| \braket{\phi_{\nuv}}{\psi} \big|^2 =
   1 - 6 \lambdaB^2 \alphaB + \Oo (\lambdaB^3)  + \Oo ( \frac{\lambdaB U_B}{V} )\;.
\end{equation}
Moreover, one has in view of (\ref{eq-1st_order_corr_GS})
\begin{equation} \label{eq-scalar_prod_3}
  \braket{\phi_\nv^{(0)}}{\psi^{(1)}}=
  \begin{cases}
    \dss
    \sqrt{\alphaB} \sum_{<i,j>, \Delta_{01}^{ij} = (\Delta \nv)_{01}} \overline{c_{\nuv+1_j-1_i,\nv}}\,\E^{\I \phi_{ij}} 
    & \text{ if $\Eint (\nv) = \Eint_0 + U_B$}\\
    0 & \text{ otherwise.}
  \end{cases}
\end{equation}

Plugging the results (\ref{eq-perturbed_eigenenergies_W}), (\ref{eq-scalar_prod_1}), (\ref{eq-scalar_prod_2}),
and (\ref{eq-scalar_prod_3}) into (\ref{eq-kk_after_spectral_decomp})
and using the fact that all Fock states with interaction energies $\Eint (\nv) = \Eint_0 + U_B$ have a single particle-hole
excitation, that is, there are of the form $\ket{\nv} = \ket{\nuv + 1_l - 1_k}$ with $k \not= l$ not necessarily nearest neighbors, 
one deduces that
\begin{equation} \label{eq-general_result_Schmidt_number}
  \Kk_{AB} (t)  = 
  8 L \alphaA \alphaB \lambdaA^2 \lambdaB^2
  \bigg[ 6
    - \sum_{k\not=l, k \;\text{or}\; l \in \{0,1\}} \Aa_{kl} \cos \Big( t \big( -\Delta^{kl}_{01} V + U_B +  E_B^{(1)} ( \nuvB + 1_l - 1_k) \big) \Big)    \bigg]
\end{equation}
 with
\begin{equation}
  \Aa_{kl} \equiv \frac{1}{\alpha_B} \big| \braket{\phi_{\nuv +1_l - 1_k}^{(0)} }{\psi^{(1)} } \big|^2
    = \Bigg| \sum_{<i,j>, \Delta_{01}^{ij}=\Delta_{01}^{kl}} c_{\nuvB + 1_j-1_i, \nuvB + 1_l - 1_k} \E^{-\I \phi_{ij}} \Bigg|^2\;.
\end{equation}
The missing error terms inside the square bracket in (\ref{eq-general_result_Schmidt_number})
are of order $\lambdaB$, $U_B/(\lambdaB V)$, $\lambdaB  J_B t$, and
$ U_B^2 t/V$.

Therefore, by inspection on  (\ref{eq-values_Delta_1_j-1_i}) and (\ref{eq-general_result_Schmidt_number}),
$\Kk_{AB} (t)$ presents oscillations with frequencies $2 V\pm U_B + \Oo (J_B)$
and  $V\pm U_B + \Oo (J_B)$ 
around the mean value $\langle \Kk_{AB} \rangle_t = 48 L \alpha_A \alpha_B \lambdaA^2 \lambdaB^2$.
In particular, Eq.~(\ref{eq-average_Schmidt_number_small_lambdaA_lambdaB}) remains valid 
when the Schmidt number is averaged over time intervals $[0, t]$ satisfying
\begin{equation} \label{eq-condition_averaging_time}
  V^{-1} \ll t \ll U_A^{-1}, (\lambdaB J_B)^{-1}, V   U_B^{-2}
\end{equation}  
(recall that Eq.~(\ref{eq-average_Schmidt_number_small_lambdaA_lambdaB}) was shown in
Sect.~\ref{eq-entanglement_B_MI_regime} to hold in the small time regime $t \ll U_B^{-1}, U_A^{-1}$ only).
Note that for large atom numbers $N_A,N_B$, all energy parameters in these inequalities must be multiplied by
the corresponding atom number, see the discussion at the end of Sec.~\ref{eq-entanglement_B_SF_regime}.
The averaged Schmidt number $\langle {\Kk_{AB}} \rangle_t$ evaluated numerically is displayed as function of $t$ in  Fig.~\ref{Fig_app3}(a).
One observes that $\langle {\Kk_{AB}} \rangle_t$ is constant in the time range $[100 V^{-1}, J_B^{-1}]$.
Note that a good agreement with the predicted value from
Eq.~(\ref{eq-average_Schmidt_number_small_lambdaA_lambdaB}) is obtained even for times
$t$ not satisfying the condition $t \ll U_A^{-1}$.

Formula (\ref{eq-general_result_Schmidt_number}) is valid  in the intermediate time regime
$t \approx J_B^{-1}$ (with $J_B \gg U_A , U_B^2/V$).
We now derive a simpler and more explicit expression for
times $0 \leq t \lesssim U_{B}^{-1} \ll J_B^{-1}$. Then the first-order correction $E_B^{(1)}$ to the eigenenergies
(\ref{eq-perturbed_eigenenergies_W}), which is of order $J_B$, can be neglected and one gets
\begin{equation}
  \Kk_{AB} (t)  =  \langle {\Kk_{AB}} \rangle_t - 8 L \alphaA \lambdaA^2 \lambdaB^2
  \sum_{\Delta = \pm 2, \pm 1}  \cos \big( t (- V \Delta + U_B ) \big) 
  \sum_{\mv, (\Delta\mv)_{01}=\Delta, E^{(0)}_B (\mv) = \EGSB + U_B} \big| \braket{\phi^{(0)}_\mv}{\psi^{(1)}_B} \big|^2\;.
\end{equation}
But for any eigenenergy $\epsilon$ of $\HB^\inter$ one has
\begin{equation}
  \sum_{\mv \in \Ss_{\Delta , \epsilon}} \big| \braket{\phi^{(0)}_\mv}{\psi^{(1)}_B} \big|^2 =  \sum_{\mv \in \Ss_{\Delta , \epsilon}} \big| \braket{\mv}{\psi^{(1)}_B} \big|^2\;.
\end{equation}
Furthermore, in view of Eq. (\ref{eq-1st_order_corr_GS}), for any $\mv \not= \nuvB$ it holds
$\braket{\mv}{\psi^{(1)}_B}  = \sqrt{\alphaB} \E^{\I \phi^B_{ij}}$ if $\mv = \nuvB + 1_j-1_i$ with $i,j$ nearest neighbors and
$\braket{\mv}{\psi^{(1)}_B}  =0$ otherwise.
This yields, using (\ref{eq-values_Delta_1_j-1_i}) again,
\begin{equation} \label{eq-time_evolution_of_K}
  \begin{array}{lcl}
      \dss \frac{1}{L} \Kk_{AB} (t) & = & 8 \alphaA \alphaB  \lambdaA^2 \lambdaB^2 \Big[ 6 - 4  \cos ( t V) \cos ( t U_B) - 2
        \cos (2 t V) \cos ( t U_B) \Big]
      \\
      & & \hspace*{7cm} \text{ if } t \ll U_A^{-1}, J_{B}^{-1}, V   U_B^{-2}\;.\end{array}
\end{equation}
Note that this formula agrees with (\ref{eq-Schmidt_number_small_lambdaA_lambdaB}) 
in the short time regime $t \ll U_B^{-1}, U_A^{-1}$. 
The time evolution of $\Kk_{AB}(t)$ calculated  numerically by exact diagonalization is compared with
our analytical result (\ref{eq-time_evolution_of_K}) in the inset of Fig.~\ref{Fig-3}(a), showing a good agreement.
{ The presence of the frequency $U_B$ in Eq.~(\ref{eq-time_evolution_of_K}), which is the particle-hole excitation energy for the species B, indicates that the entanglement 
    does not only depend on the inter-species interactions but also on the post-quench intra-species interactions.}

It is worth observing that the Schmidt number $\Kk_{AB}(t)$ 
being symmetric in $A$ and $B$, the whole calculation above remains valid by exchanging the role of $A$ and $B$.
Therefore, the time-averaged Schmidt number is also given by (\ref{eq-average_Schmidt_number_small_lambdaA_lambdaB}) for
averages up to time $t$ satisfying $V^{-1} \ll t \ll U_B^{-1}, (\lambdaA  J_A)^{-1}, V   U_A^{-2}$,
instead of (\ref{eq-condition_averaging_time}). Similarly, (\ref{eq-time_evolution_of_K}) is correct with
$U_B$ replaced by $U_A$ when $t \ll  U_{B}^{-1}, J_A^{-1}, V   U_A^{-2}$.
Furthermore, as shown in Appendix~\ref{app-B}, Eqs.~(\ref{eq-average_Schmidt_number_small_lambdaA_lambdaB})
and (\ref{eq-time_evolution_of_K}) can be easily extended to Bose-Bose mixtures in arbitrary finite lattices
$\Lambda$ in dimension $D=1$, $2$, or $3$ with periodic boundary conditions,
see Eqs. (\ref{eq-mean_Schmidt_number_arbitary_lattice}) and (\ref{eq-time_evolution_of_K_z_arbitarary_lattice}).

\subsection{Gases $A$ and $B$  in the MI and SF regimes} \label{sec-case(ii)}

Let us now determine the Schmidt number in the intermediate time regime when
$\lambdaB \gg 1$ (gas $B$  in the  SF regime).
As in the previous subsection, we rely on
the separation of energy scales (\ref{eq-separation_of_timescales}) and  determine the spectrum and eigenvectors
of $\Wop$ using perturbation theory, the perturbation being the Bose-Hubbard Hamiltonian $\Hop$ of the gas $B$.
One must diagonalize the projected Hamiltonian
\begin{equation} \label{eq-projected_Hamiltonian}
  \PiB \Hop \,\Pi_\Delta = \Pi_\Delta \Kop \,\Pi_\Delta + \Oo ( U_B)
  = \Pi_\Delta \Kop' + \Oo ( U_B)\;,
\end{equation}
where $\PiB$ is the projector (\ref{eq-projector_Pi_Delta}). In the last equality in Eq.~(\ref{eq-projected_Hamiltonian}) we  have introduced the kinetic Hamiltonian $\Kop'$ of an open  chain $\Lambda'=\{ 2, \ldots, L-1\}$ with $(L-2)$ sites,
\begin{equation} \label{eq-kinetic_hamiltonian_chain}
\Kop' = - J_B \sum_{j=2}^{L-2} ( e^{\I \phi_B} \hat{b}_{j+1}^\dagger \hat{b}_j + {\rm h.c.} ) \;.  
\end{equation}
The equality $\Pi_\Delta \Kop \,\Pi_\Delta = \Pi_\Delta \Kop'$ follows 
from the fact that $\Delta_{01}^{ij} =0$ if and only if
$i,j \notin \{ 0,1\}$, see (\ref{eq-values_Delta_1_j-1_i}).

The eigenvectors of $\Pi_\Delta \Kop'$ are given by $\ket{n_0,n_1} \ket{\phi_{\sigma,k}'}$ with
$n_1-n_0=\Delta$ and $n_0+n_1=\sigma$, $ \ket{\phi_{\sigma,k}'}$  being the
eigenvectors of $\Kop' $ for an atomic gas  with $N_\sigma \equiv \NB - \sigma$ atoms trapped in the chain $\Lambda'$, 
\ie, $\Kop' \ket{\phi_{\sigma,k}'} = w_{\sigma,k}'  \ket{\phi_{\sigma,k}'}$.
Thus the perturbative eigenvalues and eigenvectors of $\Wop$ are
\begin{eqnarray} \label{eq-perturbed_eigenenergies_Wbis}
  w_{n_0,n_1,k} & = & - V ( n_1-n_0) + w_{\sigma,k}' + \Oo (  U_B ) + \Oo \Big( \frac{ J_B^2}{V} \Big)
  \\
  \label{eq-perturbed_eigensates_Wbis}
  \ket{\phi_{n_0,n_1,k}} & = & \ket{n_0,n_1} \ket{\phi_{\sigma,k}'} + \Oo (\lambdaB^{-1}) + \Oo \Big( \frac{J_B}{V} \Big)\;.
\end{eqnarray}
Using the spectral decomposition $\E^{-\I t \Kop'}=\sum_k \E^{-\I t w_{\sigma,k}'} \ketbra{\phi_{\sigma,k}'}{\phi_{\sigma,k}'}$, we find
\begin{equation} \label{eq-expectation_Wop_SF}
\bra{\psi}  \E^{-\I t \Wop} \ket{\psi}  
\\
 = 
\sum_{n_0,n_1}  \E^{\I t V (n_1-n_0)} \braket{\psi}{n_0,n_1} \E^{-\I t \Kop ' } \braket{n_0,n_1}{\psi}
\end{equation}
up to errors of order  $\lambdaB^{-1}$, $J_B/V$, $ U_B t$, and $J_B^2 t/V$.
Note that Eq.~(\ref{eq-expectation_Wop_SF}) applies to  times satisfying
$t \ll U_B^{-1}$, in addition to (\ref{eq-intermediate_times}),
because we have neglected the interaction term $\Pi_\Delta \hat{H}^\inter$ in~(\ref{eq-projected_Hamiltonian});
it is, however,  justified to use (\ref{eq-expectation_Wop_SF}) for times $t \approx J_B^{-1}$.
To simplify we assume that $-\phi_0/2 \leq \phi_B < \phi_0/2$, so that the gas $B$ has zero angular momentum ($\ell=0$).
Replacing $\ket{\psi}$ by the SF state $\ket{\psi_\SF}$, a calculation similar to the one performed
in Sect.~\ref{eq-entanglement_B_SF_regime} leads to
\begin{equation} \label{eq-Q_expectation_inter_time_regime}
\bra{\psi}  \E^{-\I t \Wop} \ket{\psi}   =  
  \sum_{\sigma=0}^{N_B} \bigg( \frac{2 \cos (tV)}{L} \bigg)^\sigma \frac{\NB !}{\sigma! (\NB - \sigma)!} 
  \bigg( \frac{L-2}{L} \bigg)^{\NB-\sigma} R_\sigma (t)
\end{equation}
with
\begin{equation}
  R_\sigma(t) \equiv  \bra{\psi_{\SF,\sigma}'} \E^{-\I t \Kop'} \ket{\psi_{\SF,\sigma}'}\;.
\end{equation}
Here, $ \ket{\psi_{\SF,\sigma}'}$ is the SF state of a gas with $N_\sigma$ atoms
and angular momentum $\ell=0$
trapped in the ring lattice $\Lambda'_{\text{ring}}=\{2,\cdots, L-1\}$ with $(L-2)$ sites. 
Note that the identity (\ref{eq-Q_expectation_inter_time_regime}) does not hold when $\ell \not=0$ due to the different values of $\phi_0$
in the lattices $\Lambda$ and $\Lambda'_{\rm ring}$.

We now show that in the thermodynamical limit the factor
$R_\sigma(t)$
is equal to $e^{-\I t E_{\SF,\sigma}}$, where $E_{\SF,\sigma} =-2 N_\sigma J_B \cos \phi_B$ is the GS energy
of a non-interacting Bose gas with $N_\sigma$ atoms in the ring lattice $\Lambda'_{\text{ring}}$.
Actually, let us denote by  $ \Kop_{\text{1st\,q}}'$ the first quantization version of (\ref{eq-kinetic_hamiltonian_chain}) and by
$\ket{\varphi_0^{\text{ring}}}$ the GS of a single atom in the ring $\Lambda'_{\text{ring}}$.
The kinetic Hamiltonian on $\Lambda'_{\text{ring}}$ differs from $\Kop_{\text{1st\,q}}'$ by the presence of
hopping terms between sites $2$ and $L-1$, namely by $\hat{Q} =-J_B ( e^{\I \phi_B} \ketbra{2}{L-1} + {\rm h.c.})$.
We can use second-order perturbation theory to expand $\ket{\varphi_0^{\text{ring}}}$ in terms of the eigenstates $\ket{\varphi_k'}$
of $\Kop_{\text{1st\,q}}'$ and to evaluate $\bra{\varphi_0^{\text{ring}}} \E^{-\I t \Kop_{\text{1st\,q}}'} \ket{\varphi_0^{\text{ring}}}$.
Observing that $|\bra{\varphi_k'} \hat{Q} \ket{\varphi_0'}|^2$ is typically much smaller
than $L^{-1}$ for large $L$'s (the GS and first excited
states of the chain $\Lambda'$ are delocalized on the whole chain), this gives
$\bra{\varphi_0^{\text{ring}}} \E^{-\I t \Kop_{\text{1st\,q}}'} \ket{\varphi_0^{\text{ring}}}\simeq  \E^{-\I t E_0'} +  o ( L^{-1} )$, where
 $E_0'$ is the  GS energy of a single atom in the chain $\Lambda'$.
Since the SF state is the $N_\sigma$-fold tensor product of the single atom state $\ket{\varphi_0^{\text{ring}}}$
and noting that $o(L^{-1})=o( N_\sigma^{-1})$, one infers that for $N_\sigma \gg 1$,
\begin{equation} \label{eq-factor_R_sigma}
  R_\sigma(t)= \bra{\varphi_0^{\text{ring}}} \E^{-\I t \Kop_{\text{1st\,q}}'} \ket{\varphi_0^{\text{ring}}}^{N_\sigma} 
   \sim
   \E^{-\I t N_\sigma E_0'}  \sim  \E^{2\I t N_\sigma J_B \cos \phi_B}\;,
\end{equation}
where we have
approximated $E_0'$ 
by the GS energy  $E_0^{\text{ring}}= -2 J_B\cos\phi_B$ of a single atom in the ring $\Lambda'_{\text{ring}}$.

Replacing (\ref{eq-Q_expectation_inter_time_regime}) and (\ref{eq-factor_R_sigma})
into (\ref{eq-Schmidt_number_general_expression}),
one obtains in the thermodynamical limit $\NB, L \gg 1$,
\begin{equation} \label{eq-Schmidt_number_intermediate_times_SF_regime}
\dss  \frac{\Kk_{AB} (t)}{L} \simeq  4 \alpha_A \lambdaA^2 \Big( 1 - \exp \big\{ -4 \nuB \big( 1 - \cos(tV) \cos ( 2 t J_B \cos \phi_B ) \big) \big\}
  \Big)
   \; .
\end{equation}
This formula is valid for Peierls phases  $\phi_B$ satisfying  $-\phi_0/2 < \phi_B < \phi_0/2$ and
times $t$ much smaller than $ U_B^{-1}, U_A^{-1}$, and $V  J_B^{-2}$.
Notice that one recovers the result of Eq. (\ref{eq-Schmidt_number_case(ii)}) when  $t \ll J_B^{-1}$.
On the other hand, averaging $\Kk_{AB}(t)$ up to times $t$ with $J_B^{-1} \ll t \ll U_B^{-1}, U_A^{-1}, V J_B^{-2}$,
one is led to
\begin{equation} \label{eq-time_averaged_Schmidt_number_intermediate_times_SF_regime}
  \frac{\langle \Kk_{AB} \rangle_t}{L}
  \simeq   
  6 \alpha_A \lambda_A^2 \,\beta_{\nu_B,\infty}'
  \quad , \quad 
  \beta_{\nu_B,\infty}'
  =   
  \frac{2}{3}  \bigg( 1 - \lim_{k \to \infty} \int_0^{2\pi k} \frac{\D s}{2\pi k}
   \E^{ -4 \nuB ( 1 - \cos( s)  \cos (s/k) )} \bigg)\;.
\end{equation}  
This result easily follows by (i) making the change of variables $s=Vt$ in the integral between $0$ and $t$
of the exponential in (\ref{eq-Schmidt_number_intermediate_times_SF_regime});
(ii) approximating the second cosine  by $\cos (s/k)$ with $k=E[V/(2 J_B\cos\phi_B)]$
(this is justified since $t \ll V J_B^{-2}$); (iii) using the periodicity of the resulting integrand;
 (iv) taking $k = \Oo (V/J_B) \to \infty$. 

 We may summarize the results of this subsection and those of Sec.~\ref{eq-entanglement_B_SF_regime} by
 the following formula
\begin{eqnarray} \label{eq-summary_formula_Schmdit_number_B_SF}
  \frac{\langle \Kk_{AB} \rangle_t}{L}
& =  &   6 \alpha_A \lambdaA^2 
  \begin{cases}
    \beta_{\nu_B,\infty}  & \text{ if $V^{-1} \ll t \ll J_B^{-1}$}
 \\ \dss
  \beta_{\nu_B,\infty}'  
& \text{ if $J_B^{-1} \ll t \ll U_B^{-1}, U_A^{-1}, V J_B^{-2}$,}
\end{cases}
\end{eqnarray}
where $\beta_{\nuB,\infty}$ and $\beta_{\nuB,\infty}'$ are given by
(\ref{eq-beta_B_SF}) and (\ref{eq-time_averaged_Schmidt_number_intermediate_times_SF_regime});
their numerical values are given in Table~\ref{tab-1} for $\nuB=1$ and $2$.

  We compare  in Fig.~\ref{fig-new_fig}(b) formula (\ref{eq-Schmidt_number_intermediate_times_SF_regime}) with
  the Schmidt number $\Kk_{AB} (t)$ evaluated numerically.
  One sees that while Eq.~(\ref{eq-Schmidt_number_intermediate_times_SF_regime}) describes well the time evolution of
  $\Kk_{AB} (t)$ in the time range $[0, J_B^{-1}]$, at later times
  discrepancies appear; this is likely to be due to contributions of larger frequencies $1/U_B$ and $1/U_A$
  that have been neglected  in~(\ref{eq-Schmidt_number_intermediate_times_SF_regime}).
  In spite of this,
  one observes in  Fig.~\ref{Fig_app3}(b) that 
  the averaged Schmidt number $\langle {\Kk_{AB}} \rangle_t$ evaluated numerically for $L=N_A=N_B=4$
  and $\lambda_B=5$
  is constant over a wide range of time $[ J_B^{-1}, 100 U_B^{-1}]$, in analogy with what happens when $B$ is in the  MI regime.
  This indicates that the aforementioned contributions of the frequencies $1/U_A$ and $1/U_B$ average out to zero.
For Peierls phases $\phi_A=\phi_B = \pi/20$ and $\pi/3$, 
the constant value agrees reasonably well with the value predicted by 
Eq.~(\ref{eq-time_averaged_Schmidt_number_intermediate_times_SF_regime})
even for times $t$ not satisfying the conditions $t \ll U_B^{-1}$ and $t \ll U_A^{-1}$, even though a slight dependence on the phase is observed  for time $t$ larger than  $J_B^{-1}$ (in contrast, as seen in the inset, the two curves for
the phases $\pi/20$ and $\pi/3$ are not distinguishable when $t \ll J_B^{-1}$).
On the other hand, $\langle {\Kk_{AB}} \rangle_t$ is smaller for $\phi_A=\phi_B=\phi_0/2=\pi/4$, both in the short
time regime $t \lesssim J_B^{-1}$  and in the intermediate and large time regimes $J_B^{-1} \lesssim t \leq 100 U_B^{-1}$. As pointed out
above,   Eqs.~(\ref{eq-proportionality_Schmidt_number_relative_current2}), (\ref{eq-time_averaged_Schmidt_number_intermediate_times_SF_regime}), and (\ref{eq-summary_formula_Schmdit_number_B_SF}) are not valid at this phase value, since
the GS of the gas $B$ is a superposition of SF states with angular momenta $\ell=0$ and $\ell=1$. This
explains the small disagreement between our numerical and analytical results.
We observe on  Fig.~\ref{Fig_app3}(b)  that the values of $\langle {\Kk_{AB}} \rangle_t/L$ for $\phi_A=\phi_B=\phi_0/2$ are obtained 
by shifting the curve for $\phi_A=\phi_B = \pi/20$ or $\pi/3$ by some time-independent
constant of order  $0.5 \lambda_A^2$.

The fact that Eqs. (\ref{eq-Schmidt_number_intermediate_times_SF_regime}) and (\ref{eq-time_averaged_Schmidt_number_intermediate_times_SF_regime})), which have been derived assuming a large ring, describe well the
behavior of $\Kk_{AB} (t)$ at times $t \lesssim J_B^{-1}$ and its averaged value at times
$t \gg J_B^{-1}$ for a ring with $L=4$ sites is an indication that finite size effects do not play an important role.

 \section{Conclusion} \label{sec-conclusion}

We have studied the dynamics of persistent currents in a binary mixture in the presence of an interaction quench between the species.
  Assuming that one of the gas components is in the MI regime,
  we found universal relations between the persistent current of the other species, the visibility before the quench,
  and the amount of inter-species entanglement.
  In particular,  for the short time regime, the analytical expressions of the relative variation of the time-averaged  current show that it is
  proportional to the { 2-R\'enyi entropy of entanglement} and scales quadratically with the visibility before the quench.

Our results are in principle amenable to experimental realizations. In fact, 1D-ring 
lattice trapping potentials have been already achieved~\cite{Amico2014} and artificial gauge fields 
can be created  in various ways~\cite{Dalibard2011}. Interaction quench dynamics can be generated using
the ability to tune atomic interactions with Feshbach resonances~\cite{meinert,modu,Polleti}. Finally, time-of-flight
techniques for obtaining information on the atomic current are available. In particular, the winding number and
current-phase relationship can be extracted from
the spiral interference pattern obtained 
after releasing the trap in the plane of the ring and letting the BEC 
interfere with another BEC confined initially near its center~\cite{Dalibard2014, Campbell2014, Mathew_PRA2015, Amico2018}. 

{ Some open problems require further studies.
  Firstly, even though our numerical simulations indicate a saturation of the Schmidt number  at large times,
  at least after having performed a local time averaging to suppress oscillations with frequency equal to the inter-species interaction strength
  (see Fig.~\ref{Fig_app3}),
  we can not give a definite answer to the question of a possible convergence of the {  reduced density matrix of one species}
  to a stationary state.}
  In order to get more insight on the convergence properties of the out-of-equilibrium dynamics,
  one would need to perform simulations for larger
  system sizes, which is difficult with our exact diagonalization approach;
  our analytical approach is also of limited use for this problem because it applies to small or intermediate times.}  

Secondly, it would be of interest to study the $B$-current and entanglement evolutions when  the gas $A$ is in the SF regime, instead of the MI regime.
  One then expects that this gas  could have a stronger effect on the $B$-current.
  {  Note that for repulsive inter-species interactions and when both species are in the SF regime,
    it has been shown in~\cite{Penna17} that certain initial excitations lead in the  gauge free $3$-site lattice  case
    to a periodic transfer of angular momentum between the two species.}
  These issues will be  investigated in a forthcoming work.
  
  { Thirdly, it would be worth adapting our methods to 
    the case of a single impurity atom interacting with a gas of atoms of a different species, taking
   $N_A =1$ and $N_B \gg 1$ or vice versa in our model. This problem has attracted a lot of interest recently
    in relation with the
    polaron physics~\cite{Meinert_Sceince_2017,Grust_NJP_2017,Volomiev_PRA2017,Mistakidis_PRL2019,Theel_NJP_2020}.}

\vspace{0.5cm}
\noindent
{\bf\large  Acknowledgments}
D.S. acknowledges support from  the Fondecyt project N$^0$ 1190134 and
the  Vicerector\'{\i}a de Investigaci\'on y Desarollo de  la Universidad de Concepci\'on, proyecto VRID 218.013.045-1.OIN,
and L. M-M. acknowleges support  from  the Fondecyt project N$^0$ 1190629 and 
the Vicerector\'{\i}a de Investigaci\'on de la Pontificia
Universidad Cat\'olica de Chile, proyecto Puente No 03/2019.

\vspace{1cm}

\appendix
\renewcommand{\thesection}{\Alph{section}}
\numberwithin{equation}{section}
\setcounter{equation}{0}


\section{Visibility of a BEC trapped in a 1D-ring lattice potential}
\label{supplementary_material}

In this appendix we determine the visibility of the interference fringes of a single BEC 
trapped in a 1D-ring lattice potential in the presence
of an artificial gauge field.
As mentioned in the main text, the interference pattern is obtained after releasing the potential trap  and letting the atomic gas expand freely, 
thus it is directly related to the momentum distribution
\begin{equation} \label{eq-def_momentum_distrib}
  S_\phi (q) = \sum_{i,j=0}^{L-1} \E^{\I q (i-j)} \bra{\psi_\phi} \bop_i^\dagger \bop_j \ket{\psi_\phi}\;,
\end{equation}
where $\ket{\psi_\phi}$ is the ground state (GS) of the Bose-Hubbard Hamiltonian $\Hphi$ of a single species
(Eq. (1) in the main text),
$\phi$ is the Peierls phase,
$\bop_i^\dagger$ and $\bop_j$ are the creation and annihilation operators of a bosonic atom at sites $i$ and $j$,
and $L$ is the number of lattice sites.
The visibility is defined as 
\begin{equation} \label{eq-def_visibility}
  {\cal V}_\phi = \frac{S_{\max}-S_{\min}}{S_{\max}+S_{\min}}\;,
\end{equation}
$S_\Max$ and $S_\Min$ being the maximum and minimum of $S_\phi(q)$.
By gauge invariance,  ${\cal V}_\phi$ is periodic in $\phi$ with period $\phi_0=2\pi/L$. Actually, 
the Hamiltonians $\hat{H}_{\phi+\phi_0}$ and $\Hphi$ and their GSs are related
by a gauge transformation,
e.g. $\ket{\psi_{\phi + \phi_0}} = \E^{ \I \phi_0 \hat{X}}  \ket{\psi_{\phi}}$ with $\hat{X} = \sum_j j \bop_j^\dagger \bop_j$,
implying that $S_{\phi+\phi_0} (q) =S_{\phi} ( q -\phi_0)$.

We determine  analytically the momentum distribution and visibility when the Bose gas is in the Mott-insulator (MI) regime, \ie, for small energy ratio $\lambda = J/U$.
We rely on standard perturbation theory, the perturbation being the kinetic term
\begin{equation}
  \Kphi = - J \sum_{j=0}^{L-1} (  \E^{\I \phi} \bop_{j+1}^\dagger \bop_j + {\rm h.c.} )\;.
\end{equation}
As in the main text, the Fock states diagonalizing $\hat{H}^\inter = \Hphi - \Kphi$ are denoted by  $\ket{\nv}= \ket{n_0,\ldots, n_{L-1}}$,
$E^{(0)}(\nv)$ are the corresponding eigenenergies, the filling factor $\nu=N/L$ is assumed to be an integer,
and $\ket{\psi_\MI}=\ket{\nuv}=\ket{\nu,\ldots, \nu}$ is the MI state (GS of $\hat{H}^\inter$ with energy $E_0^{(0)}= E (\nuv)$).
For small $\lambda$, the GS of the BEC reads
\begin{equation} \label{eq-pertubrative_expansion_GS-app}
  \ket{\psi_\phi} = \ket{\psi_\MI} + \lambda \ket{\psi^{(1)}_\phi} + \lambda^2 \ket{\psi^{(2)}_\phi} + \Oo ( \lambda^3)
\end{equation}
with  the first and second-order corrections
\begin{eqnarray} \label{eq-1st_order_corr_GS_app-A}
  \lambda \ket{\psi^{(1)}_\phi} & = &  - \frac{1}{U} \hat{K}_\phi \ket{\psi_\MI}
  \;=\; \lambda \sum_{j=0}^{L-1} \big(  \E^{\I \phi} \bop_{j+1}^\dagger \bop_j + \adj \bigr) \ket{\psi_\MI}
\\ \label{eq-2nd_order_corr_GS}
\lambda^2 \ket{\psi^{(2)}_\phi} & = &
\sum_{\mv, \pv \not= \nuv} \frac{\bra{\mv} \Kphi \ket{\pv} \bra{\pv} \Kphi \ket{\psi_\MI}}{( E_0^{(0)} - E^{(0)} ( \mv)) ( E_0^{(0)} - E^{(0)} ( \pv))} \ket{\mv}
 - \onehalf \sum_{\mv \not= \nuv} \frac{ | \bra{\mv} \Kphi \ket{\psi_\MI} |^2}{( E_0^{(0)} - E^{(0)} ( \mv))^2} \ket{\psi_\MI}\;.
\end{eqnarray}
Here, we have used $\bra{\psi_\MI} \Kphi \ket{\psi_\MI} = 0$ and the fact that the energy to create a particle-hole excitation
$\propto \bop_i^\dagger \bop_j \ket{\psi_\MI}$ is
equal to
\begin{equation} \label{eq-particle_hole_energy}
E^{(0)} ( \nuv + 1_i - 1_j ) - E^{(0)}_0=  U\quad, \quad i \not= j\;,
\end{equation}
where $1_j=(0,\ldots, 1, 0, \ldots , 0)$ is the vector having a single nonzero component  equal to unity at site $j$. 
Noting that $\bra{\mv} \hat{K}_\phi \ket{\psi_\MI}=0$ if $E^{(0)}(\mv) \not= E_0^{(0)} + U$, one deduces from (\ref{eq-2nd_order_corr_GS}) that
\begin{equation} \label{eq-scal_prod_zero_and_2nd_order_WF}
  \lambda^2 \braket{\psi_\MI}{\psi_\phi^{(2)}} = - \frac{1}{2U^2} \bra{\psi_\MI} \hat{K}_\phi^2 \ket{\psi_\MI}
  = -\frac{\lambda^2}{2} \big\| \psi^{(1)}_\phi \big\|^2 = - \nu(\nu+1)\lambda^2  L\;.
\end{equation}
Substituting (\ref{eq-pertubrative_expansion_GS-app}) into (\ref{eq-def_momentum_distrib}), the momentum distribution reads
\begin{eqnarray} \label{eq-perturbative_momentum_distrib1}
  \nonumber
  S_\phi (q) & = & \sum_{i,j=0}^{L-1} \E^{\I q (i-j)}   \bra{\psi_\MI} \bop_i^\dagger \bop_j \ket{\psi_\MI}
  + 2 \lambda \re \bigg\{  \sum_{i,j=0}^{L-1} \E^{\I q (i-j)}  \bra{\psi_\MI} \bop_i^\dagger \bop_j \ket{\psi^{(1)}_\phi} \bigg\}
  \\
  & & 
  + 2 \lambda^2 \re \bigg\{  \sum_{i,j=0}^{L-1} \E^{\I q (i-j)}  \bra{\psi_\MI} \bop_i^\dagger \bop_j \ket{\psi^{(2)}_\phi} \bigg\}
   + \lambda^2\sum_{i,j=0}^{L-1} \E^{\I q (i-j)}  \bra{\psi^{(1)}_\phi} \bop_i^\dagger \bop_j \ket{\psi^{(1)}_\phi}
   + \Oo ( \lambda^3 )\;.
\end{eqnarray}   
A simple calculation yields
\begin{eqnarray} \label{eq-matrix elements_tunelling_term}
\nonumber
  \bra{\psi_\MI} \bop_i^\dagger \bop_j \ket{\psi_\MI} & =  & \nu \delta_{ij} 
\\ \nonumber
\bra{\psi_\MI} \bop_i^\dagger \bop_j \ket{\psi^{(1)}_\phi} & = &
\nu ( \nu+1) (1 -\delta_{i,j})  \sum_{l=0}^{L-1} \Big( \E^{\I \phi}  \delta_{i,l} \delta_{j,l+1} + \E^{-\I \phi} \delta_{i,l+1} \delta_{j,l}  \Big) 
\\ 
\bra{\psi_\MI} \bop_i^\dagger \bop_j \ket{\psi^{(2)}_\phi} & = &
\nu ( \nu+1) \bigg[ ( 1 - \delta_{i,j} ) \sum_{l=0}^{L-1} \Big( (\nu+1) \E^{2 \I \phi} \delta_{i,l} \delta_{j,l+2} +  \nu \E^{2 \I \phi} \delta_{i,l-1} \delta_{j,l+1}
 \\ \nonumber
 & & + (\nu+1) \E^{-2\I \phi} \delta_{i,l+1} \delta_{j,l-1} + \nu \E^{-2 \I \phi} \delta_{i,l+2} \delta_{j,l} \Big) - \nu L \delta_{i,j}  \bigg]
\\ \nonumber
\bra{\psi^{(1)}_\phi} \bop_i^\dagger \bop_j \ket{\psi^{(1)}_\phi} & = & \bra{\psi_\MI} \bop_i^\dagger \bop_j \ket{\psi^{(2)}_\phi} + 3 \nu^2 ( \nu+1) L \delta_{i,j}\;,
\end{eqnarray}
where, according to the periodic boundary conditions, $\delta_{j,L}$ must be identified with $\delta_{j,0}$, etc...
Note that the expressions (\ref{eq-matrix elements_tunelling_term}) for $i=j$ immediately follow
from $b_i^\dagger b_i \ket{\psi_\MI}= \nu \ket{\psi_\MI}$, $\braket{\psi_\MI}{\psi_\phi^{(1)}}=0$, and (\ref{eq-scal_prod_zero_and_2nd_order_WF}).
Plugging (\ref{eq-matrix elements_tunelling_term}) into (\ref{eq-perturbative_momentum_distrib1}), one finds
\begin{equation} \label{eq-momentum_distrib}
  \begin{array}{lcl} 
    S_\phi (q) & = & \dss
    L \nu \bigg\{ 1
    + 4 (\nu + 1 ) \lambda \Big[ \Big( 1 - \frac{1}{L} \Big) \cos ( q - \phi) + \frac{1}{L} \cos \big( q (L-1) + \phi \big) \Big]
      \\
      & & \dss
      +  6 (\nu + 1) (2 \nu+1) \lambda^2 \Big[ \Big( 1 - \frac{2}{L} \Big) \cos \big( 2 ( q - \phi) \big) + \frac{2}{L} \cos \big( q (L-2) + 2 \phi \big) \Big]
     + \Oo( \lambda^3) \bigg\}\;.
  \end{array}
\end{equation}  
One easily checks that this expression fulfills  $S_{\phi+\phi_0}(q)= S_\phi ( q - \phi_0)$ (gauge invariance) and $S_{-\phi}(q) = S_{\phi}(-q)$.
As a consequence, the visibility (\ref{eq-def_visibility}) satisfies
${\cal V}_{\phi_0 \mp \phi} = {\cal V}_\phi$ and it is enough to let $\phi$ varies between  $0$ and $\phi_0/2=\pi/L$. 

A non-perturbative expression of $S(q)$ has been determined in Refs.~\cite{Gerbier05,Sengupta2005} for infinite lattices
in the absence of gauge field. Expanding the result of the first reference for small $\lambda$'s, one recovers Eq.(\ref{eq-momentum_distrib}) in the
special case $\phi=0$ and for an infinite ring ($L \to \infty$), up to an irrelevant constant term.
However, as stressed in the main text, we are interested in this article in finite rings with a non-zero gauge field.

We now determine the extrema of $S_\phi(q)$. We assume $L \geq 3$ and $\lambda \geq 0$. One has
\begin{eqnarray} \label{eq-derivative_momentum_distrib}
  \nonumber
  \frac{\partial S_\phi}{\partial q} & = &
   - 4 \lambda L \nu ( \nu+1)  \bigg\{ \Big( 1 - \frac{1}{L} \Big) \Big[  \sin ( q - \phi) + \sin \big( q (L-1) + \phi \big) \Big]
  \\
  & &
  + 3 \lambda (2\nu+1) \Big( 1 - \frac{2}{L} \Big) \Big[  \sin \big( 2( q - \phi)\big) + \sin \big( q (L-2) + 2 \phi \big) \Big]
 + \Oo (\lambda^3) \bigg\}\;.
\end{eqnarray}
We first obtain the extrema of $S_\phi$ to lowest order in $\lambda$.
Neglecting the terms of order $\lambda^2$, $\partial S_\phi/\partial q$ vanishes when
$q-\phi = - ( q(L-1)+\phi) + 2 m \pi$ or $q-\phi =  q(L-1)+\phi - (2 m +1) \pi$, \ie, 
\begin{equation}
  q= q_{m} = \frac{2 m \pi}{L}
  \quad \text{ or } \quad q=q_{m}' = \frac{(2 m +1)\pi - 2 \phi}{L-2}
\end{equation}
with $m$ an arbitrary integer.
At these extremal points, $S_\phi(q)$ takes the values
\begin{eqnarray}
  \nonumber
  S_\phi ( q_{m} ) & = & L \nu \Bigl\{ 1 + 4 (\nu+1) \lambda \cos \big(  q_{m} - \phi \big) + \Oo (\lambda^2 ) \Bigl\}
  \\
  S_\phi ( q_{m}' ) & = & L \nu \Bigl\{ 1 + 4 (\nu+1) \Bigl( 1 - \frac{2}{L} \Big) \lambda \cos \big(  q_{m}' - \phi \big) + \Oo (\lambda^2) \Bigl\}\;.
\end{eqnarray}
To determine which of the extremal  points $q_m, q_{m}'$ corresponds to the global maximum (minimum), we are left with the task of maximizing (minimizing)
$C_{m} (\phi)= \cos (  q_{m} - \phi )$ and $C_{m}' (\phi)= \cos (  q_{m}' - \phi )$ over all $m\in \{0,\ldots, L-1\}$ and $m \in \{0,\ldots, L-3\}$, respectively.

Under the assumption $0 \leq \phi \leq \pi/L$, one has $\max_{m} \{ C_m (\phi) \} = C_0 (\phi) = \cos \phi$. 
Furthermore, it follows from the bound $1 - 2/L < \cos (\pi/L)$ that  for any $m$,
\begin{equation} \label{eq-bound_C'}
- \cos \Big( \frac{\pi}{L} \Big) <  \Big( 1-\frac{2}{L} \Big) C_{m}' (\phi) < \cos \Big( \frac{\pi}{L} \Big) \leq \cos \phi\;.
\end{equation}
Thus, to lowest order in $\lambda$, the maximum of $S_\phi(q)$ is reached for $q=q_0=0$ modulo $2\pi$. Since
the term of order $\lambda^2$  in the derivative in Eq.(\ref{eq-derivative_momentum_distrib}) also vanishes for $q=q_0$,
the exact momenta at which $S_\phi(q)$ reaches its maximum differ from $q_0, q_0 \pm 2\pi, \ldots $ by at most $\Oo (\lambda^2)$. The value of the maximum is
\begin{equation} \label{eq-S_min_L_even}
S_\Max =  L \nu \Bigl\{ 1 + 4 (\nu+1)  \lambda  \cos \phi + 6 (\nu+1) ( 2 \nu+1) \lambda^2 \cos (2 \phi) + \Oo (\lambda^3 ) \Bigl\} \;.
\end{equation}

Similarly,  $\min_m \{ C_m(\phi)  \}= C_{L/2} (\phi) = - \cos \phi$ if the number of sites $L$ is even and
$\min_m \{ C_m(\phi)  \}=C_{(L+1)/2} (\phi) = - \cos (\pi/L-\phi)$ if $L$ is odd. By (\ref{eq-bound_C'}), 
the minimum of $S_\phi(q)$ is, to lowest order in $\lambda$, reached for $q=q_{L/2}= \pi$ modulo $2\pi$ in the first case and
for $q= q_{(L+1)/2}=\pi+\pi/L$ modulo $2\pi$ in the last case. Again, the term of order $\lambda^2$ in the derivative in
(\ref{eq-derivative_momentum_distrib}) vanishes at these momenta, thus these formulas give the locations of the momenta minimizing $S_\phi(q)$
up to corrections of order $\Oo (\lambda^2)$.
The value of the minimum is 
\begin{equation} \label{eq-S_max_L_even}
  S_\Min =
  \begin{cases}
    L \nu \Bigl\{ 1 - 4 (\nu+1) \lambda  \cos \phi  + 6 (\nu+1) ( 2 \nu+1) \lambda^2 \cos (2 \phi) + \Oo (\lambda^3 ) \Bigl\} & \text{ if $L$ is even}
    \\[2mm]
    L \nu \Bigl\{ 1 - 4 (\nu+1) \lambda  \cos \big( \frac{\pi}{L} - \phi \big)  + 6 (\nu+1) ( 2 \nu+1) \lambda^2 \cos  \big( \frac{2\pi}{L} - 2\phi \big)  + \Oo (\lambda^3 ) \Bigl\} & \text{ if $L$ is odd.}
  \end{cases}  
\end{equation}
We note that when  $\phi=\pi/L$   (respectively $\phi=0$ and $L$ is odd), $S_\phi(q)$ reaches its maximum (minimum)
at two phase points  in the interval $[0,2\pi]$, namely at $q_0=0$ and $q_1=2\pi/L$
(respectively at $q_{(L+1)/2}=\pi+\pi/L$ and $q_{(L-1)/2}=\pi-\pi/L$),
up to errors $\Oo (\lambda^2)$.

Even though we are considering a gas in the MI regime, it is convenient
to introduce the angular momentum of the superfluid (SF) state,  $\ell =  E [\phi /\phi_0+ 1/2]$,
where $E$ denotes the integer part,  and similarly let $k = E [\phi/\phi_0]$.
The introduction of the integers $\ell$ and $k$ helps to write the result in a concise form
for arbitrary phases $\phi \in \real$,
taking into account the periodicity and symmetry properties $ {\cal V}_{\pm \phi + \phi_0} =  {\cal V}_\phi$.
We conclude from (\ref{eq-S_min_L_even}) and (\ref{eq-S_max_L_even}) that if $L$ is even,
$L\geq 4$, the visibility is given by
\begin{equation} \label{eq-visibility_L_even}
    {\cal V}_\phi =
    4 (\nu+1) \lambda 
    \cos \big( \phi -  \ell  \phi_0  \big)
  + \Oo ( \lambda^3) 
  \quad , \quad \text{$L$ even.}
\end{equation}
If $L$ is odd, $L\geq 3$, it is given by
\begin{equation} \label{eq-visibility_L_odd}
  \begin{array}{lcl}  
    {\cal V}_\phi & = &  \dss 2 (\nu+1) \lambda 
    \Big[ \cos \Big( \frac{\phi_0}{2} - \phi + k \phi_0 \Big) + \cos \big( \phi - \ell \phi_0   \big)
    \Big] \times
   \\  
   & & \dss 
   \bigg\{ 1 - (4 \nu +1) \lambda  \Big[ \cos \Big( \frac{\phi_0}{2} - \phi + k \phi_0 \Big) - \cos \big( \phi - \ell \phi_0   \big) \Big]
      \bigg\}
  + \Oo ( \lambda^3) \quad , \quad \text{$L$ odd.}
  \end{array}
\end{equation}
%
Note that in the absence of gauge field, the visibility of ring lattices with even numbers of sites $L$ coincides  with the visibility of an infinite chain, ${\cal V}_{\phi=0} = 4 (\nu+1) \lambda  + \Oo(\lambda^3)$. The latter has been determined in Ref.~\cite{Gerbier05}.
In contrast, for finite odd numbers of sites ${\cal V}_{\phi=0} = 2 (\nu+1) ( 1+ \cos (\pi/L)) \lambda  + {\cal O}(\lambda^2)$
differs from the aforementioned gauge-free result of Ref.~\cite{Gerbier05},
which holds  in the large $L$ limit only.

It is easy to show from Eqs. (\ref{eq-visibility_L_even}) and (\ref{eq-visibility_L_odd}) that
${\cal V}_\phi$ is minimum at  half-integer values of $\phi_0$.
If $L$ is even, this follows from the fact that $\cos ( \phi - \ell \phi_0 )$ is minimum at such phase values. 
If $L$ is odd, the first-order contribution to the visibility in (\ref{eq-visibility_L_odd}) is  minimum for
$\phi= m \phi_0$ and $\phi=(m+1/2)\phi_0$, with $m \in \integer$.
By comparing the second-order corrections,
one sees that ${\cal V}_\phi$ reaches its minimum for the latter phase point.

\begin{figure}
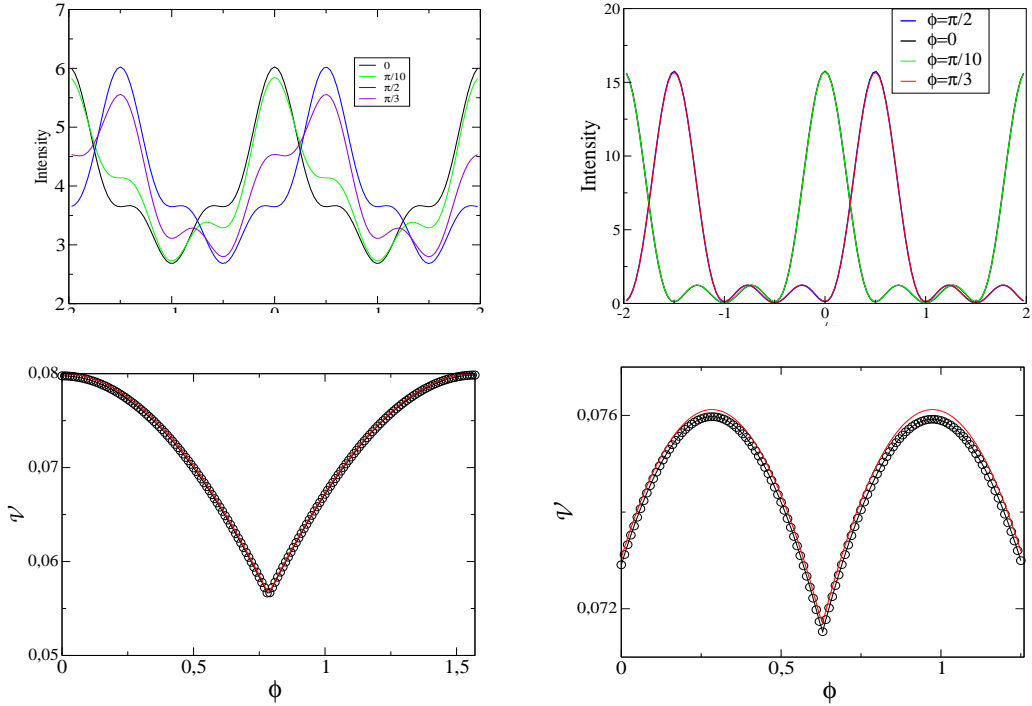

    \begin{center}
      \hspace*{3mm} \includegraphics[width=6cm, height=4.5cm, angle=0]{Fig7a.eps}
      \hspace*{10mm}  \includegraphics[width=6cm, height=4.5cm, angle=0]{Fig7b.eps}

      \vspace{3mm}
      \includegraphics[width=6.3cm,height=4.5cm,angle=0]{Fig7c.eps}
      \hspace*{7mm} \includegraphics[width=6.3cm,height=4.5cm,angle=0]{Fig7d.eps}
    \end{center}
    \caption{{\bf Upper panels:} momentum distribution $S_\phi (q)$ (in arbitrary units) as function of $q$ for a single BEC in the MI regime (left panel, $\lambda=J/U= 0.05$) and in the SF regime (right panel, $\lambda=1$), from numerical calculations;
       $N=4$ atoms are trapped
      in a 1D-ring lattice potential with $L=4$ sites  and Peierls phases $\phi=0,\pi/10,\pi/3$, and $\pi/2$.
      {\bf Lower panels:} visibility  of a Bose gas in the MI regime
      as function of the Peierls phase   
      for $N=L=4$ (left panel) and $N=L=5$ (right panel), with  $\lambda=J/U=0.01$.
      The black circles correspond to numerical calculations and the red plain curves to
      Eqs.~(\ref{eq-visibility_L_even}) and
      (\ref{eq-visibility_L_odd}).
    }
    \label{Fig_app1}
\end{figure}

The visibility is displayed as function of $\phi$ in   Fig.~\ref{Fig_app1} (lower panels) for ring lattices with
$L=4$ and $5$ sites and $\lambda=0.01$. The values obtained from Eqs. (\ref{eq-visibility_L_even})-(\ref{eq-visibility_L_odd}) agree
well with those obtained by locating numerically the extrema of the
momentum distribution. The latter is obtained numerically via an exact diagonalization of the Bose-Hubbard Hamiltonian.
We have also calculated numerically the momentum distribution and visibility for a gas 
in the SF regime ($\lambda=1$) and in the transition regime ($\lambda \simeq 0.2$).
The plots of the visibility in the transition regime are shown in Fig.~\ref{fig-1}  in the main text.
The plots of the  momentum distribution in the SF and MI regimes are displayed 
in Fig.~\ref{Fig_app1} (upper panels).
We find from our numerical results that  the locations of the global extrema of    
$S_\phi(q)$ are the same in both regimes. For instance, it is seen in Fig.~\ref{Fig_app1} that
$S_\phi(q)$ reaches its maximum at $q=\ell \phi_0$ modulo $2\pi$
in both regimes.

\section{Schmidt number as an expectation of a local time-evolution operator}
 \label{app-B}

In this appendix we derive formula (\ref{eq-Schmidt_number_general_expression}) for the shifted Schmidt number $\Kk_{AB}(t)$
quantifying the entanglement between the two gases
after interspecies interactions have been switched on.

Our analysis applies here to trapping potentials forming an arbitrary finite
lattice $\Lambda$  with $L$ sites in dimension $D=1$, $2$, or $3$, with periodic boundary conditions.
We denote by $z$ the number of nearest neighbors of a given site $i \in \Lambda$.
 For instance, $z=2$ for a $1D$ ring lattice and $z=4$ for a $2D$ square lattice.
 As we will show below, the  Schmidt number turns out to depend
 on the geometry of the lattice through $L$ and $z$ only.
 The kinetic part $\KB$ of the Bose-Hubbard Hamiltonian of the gas $B$ is given by Eq.~(\ref{eq-kinetic_part_BH}) in the main text,
 where the sum runs over all pairs of nearest neighbor sites $i,j \in \Lambda$ and the
phases $\phi_{ij} \in \{  \phi_B, - \phi_B\}$ are translation-invariant, invariant under rotation over the origin, and satisfy
$\phi_{ji} = - \phi_{ij}$ for any neighboring sites $i$ and $j$, in order to insure that $\KB$ be self-adjoint.
A similar expression holds for the kinetic Hamiltonian $\KA$ of the gas $A$.

The wavefunction of the Bose mixture  at time $t$ is
$\ket{\psi_{AB} (t)} = \E^{-\I t \HAB} \ket{\psi_A} \otimes \ket{\psi_B}$,
where  $\HAB$ is the Hamiltonian after interactions have been turned on [see Eq.~(2) in the main text] and
$\ket{\psi_A}$ and $\ket{\psi_B}$ are the GSs of the gases $A$ and $B$.
We assume
that the gas $A$ is in the MI regime ($\lambda_A \ll 1$), so that $\ket{\psi_A}$ has
the form given in (\ref{eq-pertubrative_expansion_GS-app}).
Expanding in powers of $\lambda_A$, the reduced density matrix of the gas $A$ reads
\begin{equation} \label{eq-perturbative_expansion_rhoA}
  \hat{\rho}_A (t) = \tr_B [ \ketbra{\psi_{AB} (t)}{\psi_{AB} (t)} ]
  = \hat{\rho}_A^{(0)} (t) + \lambda_A \hat{\rho}_A^{(1)} (t) + \lambda_A^2 \hat{\rho}_A^{(2)} (t)  + \Oo (\lambda_A^3 )
\end{equation}
with
\begin{eqnarray} \label{eq-def_rho^(i)}
  \nonumber
   \hat{\rho}_A^{(0)} (t)
 & =  &
   \tr_B \big[ \E^{-\I t \HAB} \ketbra{\psi_\MI}{\psi_\MI} \otimes \ketbra{\psi_B}{\psi_B} \, \E^{\I t \HAB} \big]
   \\
     \hat{\rho}_A^{(1)} (t)
 & =  &
   \tr_B \big[ \E^{-\I t \HAB}
     \big( \ketbra{\psi_A^{(1)}}{\psi_\MI} + \ketbra{\psi_\MI}{\psi_A^{(1)}} \big) \otimes \ketbra{\psi_B}{\psi_B}\,
       \E^{\I t \HAB} \big]
   \\  \nonumber
    \hat{\rho}_A^{(2)} (t)
 & =  &
    \tr_B \big[ \E^{-\I t \HAB}
      \big( \ketbra{\psi_A^{(2)}}{\psi_\MI} + \ketbra{\psi_\MI}{\psi_A^{(2)}} + \ketbra{\psi_A^{(1)}}{\psi_A^{(1)}} \big) \otimes \ketbra{\psi_B}{\psi_B}\,
        \E^{\I t \HAB} \big]\;.
\end{eqnarray}

For times $t$ satisfying $t \ll U_A^{-1}$, the kinetic Hamiltonian $\KA$ of the gas $A$ can be dropped out, so that
\begin{equation} \label{eq-approx_totoal_Hamilotnian}
  \HAB  \simeq \HA^\inter + \underbrace{\HB^\inter + \KB}_{= \HB}  + \HAB^{\inter}\quad , \quad
  \HAB^\inter = - V \nopvA \cdot \nopvB = - V \sum_{j \in \Lambda} \nopAj \nopBj 
\end{equation}
(for simplicity we do not write the identity operators explicitly,
\eg, $\HA^{\rm int}$ stands for $\HA^{\rm int} \otimes \identity_B$).
Making this approximation in (\ref{eq-def_rho^(i)}) and 
using $\HA^{\rm int} \ket{\psi_\MI} = \EGSA \ket{\psi_\MI}$ and $\HAB^{\rm int} \ket{\psi_\MI} = - V \nu_A \NB \ket{\psi_\MI}$, one easily finds that $\hat{\rho}_A^{(0)} (t)$ is independent of time and given by
\begin{equation} \label{eq-rhoA_0}
  \hat{\rho}_A^{(0)} (t) = \ketbra{\psi_\MI}{\psi_\MI}\;.
\end{equation}  
Consequently, plugging (\ref{eq-perturbative_expansion_rhoA}) and (\ref{eq-rhoA_0}) into Eq.~(\ref{eq-def_Schmidt_number}) in the main text,
\begin{equation} \label{eq-K_0}
  (\Kk_{AB} (t)+1)^{-1}
    =  1 + 2 \lambda_A \bra{\psi_\MI} \hat{\rho}_A^{(1)} (t)  \ket{\psi_\MI}
  +  2 \lambdaA^2 \bra{\psi_\MI} \hat{\rho}_A^{(2)} (t) \ket{\psi_\MI} + \lambda_A^2 \tr \big[ \hat{\rho}_A^{(1)} (t)^2 \big] \big) + \Oo (\lambda_A^3 ) \;. 
\end{equation}
The linear term in $\lambda_A$ vanishes. In fact,
$\bra{\psi_\MI}  \hat{\rho}_A^{(1)} (t)  \ket{\psi_\MI} = 2 \re \braket{\psi_\MI}{\psi_A^{(1)}}$ by (\ref{eq-def_rho^(i)})  and (\ref{eq-approx_totoal_Hamilotnian}),
and $\braket{\psi_\MI}{\psi_A^{(1)}} = 0$, see~(\ref{eq-1st_order_corr_GS_app-A}).
Generalizing (\ref{eq-scal_prod_zero_and_2nd_order_WF}) to the case of a general lattice $\Lambda$, one has
\begin{equation} \label{eq-scalar_product_psiMI_psi^(2)}
  \lambdaA^2 \bra{\psi_\MI}  \hat{\rho}_A^{(2)} (t)  \ket{\psi_\MI}
  = 2 \lambdaA^2 \re \braket{\psi_\MI}{\psi_A^{(2)}} 
= - \lambda_A^2 \alphaA L z
\end{equation}
with $\alphaA = \nuA ( \nuA+1)$.

Therefore, the only time-dependent contribution to $\Kk_{AB} (t)$ is due to the trace of $\hat{\rho}_A^{(1)} (t)^2$ in (\ref{eq-K_0}),
which can be estimated as follows.
Let us set $\hat{A}^{(1)}= \ketbra{\psi_A^{(1)}}{\psi_\MI} + \ketbra{\psi_\MI}{\psi_A^{(1)}}$.
Since $\ket{\psi_A^{(1)}} = - J_A^{-1} \KA \ket{\psi_\MI}$,
see~\ref{eq-1st_order_corr_GS_app-A}, it holds
\begin{equation}
  \big| \bra{\nv_A} \hat{A}^{(1)} \ket{\mv_A} \big|^2 = \alpha_A \sum_{<i,j>} \delta_{\nvA, \nuv_A + 1_j - 1_i}\, \delta_{\mvA,\nuvA}
  +  ( \nv_A \leftrightarrow \mv_A ) 
\end{equation}
with  $\ket{\nuv_A + 1_j - 1_i}$ the Fock state differing from $\ket{\psi_\MI}$ by
the presence of an extra particle at site $j$ and a hole at site $i$.
By evaluating the trace in the Fock basis $\{ \ket{\nvA} \} $
diagonalizing both $\nopA$ and $\HA^{\rm int}$, 
one gets from (\ref{eq-def_rho^(i)}) and (\ref{eq-approx_totoal_Hamilotnian})
\begin{eqnarray}
\nonumber  \tr_A \big[ \hat{\rho}_A^{(1)} (t)^2 \big]
  & = & 
 \sum_{\nvA, \mvA} \big| \bra{\nvA} \hat{\rho}_A^{(1)} (t) \ket{\mvA} \big|^2
 \\
\nonumber 
 & = & \sum_{\nvA, \mvA} \big| \bra{\nvA} \hat{A}^{(1)} \ket{\mvA}
\bra{\psi_B} \E^{\I t ( \HB - V \mvA \cdot \nopvB )} \E^{-\I t ( \HB - V \nvA \cdot \nopvB )}  \ket{\psi_B}  \big|^2
\\
& = &
2 \alphaA \sum_{<i,j>} \big| \bra{\psi_B} \E^{\I t  (\HB - V \nuA \NopB )} \E^{-\I t ( \HB - V \nuA \NopB - V ( \hat{n}_j^B - \hat{n}_i^B ))}  \ket{\psi_B}  \big|^2
\;.
\end{eqnarray}
Thanks to the commutation of the Bose-Hubbard Hamiltonian $\HB$ with the total number operator $\NopB$ of the $B$-atoms
and since $\ket{\psi_B}$ is an eigenstate of $\HB$, it holds
\begin{equation}
\big| \bra{\psi_B} \E^{\I t ( \HB - V \nuA \NopB )} \E^{-\I t ( \HB- V \nuA \NopB  - V ( \nopBj - \nopBi ) )} \ket{\psi_B} \big|
= \big| \bra{\psi_B}  \E^{-\I t ( \HB - V ( \nopBj-\nopBi))} \ket{\psi_B} \big|\; .
\end{equation}

Now, it follows from the translation-invariance of $\HB$ and its invariance under rotations around the origin that
\begin{equation}
  \bra{\psi_B}  \E^{-\I t ( \HB - V ( \nopBj-\nopBi))} \ket{\psi_B}
  =  \bra{\psi_B}  \E^{-\I t ( \HB - V ( \nopBone-\nopBzero))} \ket{\psi_B}
  \quad {\text{if $\;i,j$ are nearest neighbors,}}
\end{equation}
where $1$ denotes an arbitrary nearest neighbor site to the origin $0$. 
Collecting the  above results, one finds
\begin{equation}  \label{eq-expression_K_in_terms_expect_evol_op}
\Kk_{A B} (t) = 2 L z \alpha_A \lambdaA^2
\Big( 1 - \big| \bra{\psi_B} \E^{-\I t ( \HB - V ( \nopBone-\nopBzero))} \ket{\psi_B} \big|^2 \Big) + \Oo (\lambda_A^3)\;.
\end{equation}
For a 1D-ring lattice, this formula reduces to Eq.~(\ref{eq-Schmidt_number_general_expression} ) in the main text.

Recall that Eq.~(\ref{eq-expression_K_in_terms_expect_evol_op}) has been obtained by neglecting the kinetic Hamiltonian $\KA$ of the gas $A$ in the dynamics following the interaction quench.
Since $\Kk_{AB} (t) $ is of order $\lambdaA^2$, this is justified  {\it a posteriori} provided that the product $J_A t$ is much smaller than $\lambdaA$.
In fact,
by including in our calculations the corrections due to  $\KA$ estimated from time-dependent perturbation theory, one finds that
such corrections are of order $(J_A t)^2$ and $\lambdaA (J_A t)$ (all terms proportional to $J_A$ vanish due to $\bra{\psi_\MI} \KA \ket{\psi_\MI}=0$).
Hence the expression in the RHS of (\ref{eq-expression_K_in_terms_expect_evol_op}) approximates well $\Kk_{AB} (t) $ for times satisfying
$t \ll U_A^{-1}$, as one may suspect from the fact that neither $J_A$ nor $U_A$ appear in this expression.

Let us note that it is straightforward to extend the arguments of Sect.~\ref{sec-Schmidt_number_case(i)} to the case of an arbitrary lattice $\Lambda$.
One finds that when the gas $B$ is in the MI regime, the time-averaged shifted Schmidt number reads 
\begin{equation} \label{eq-mean_Schmidt_number_arbitary_lattice}
 \langle{\Kk_{AB}} \rangle_t = 8 L \alphaA \alphaB z (2z-1) \lambdaA^2 \lambdaB^2
\end{equation}  
for times satisfying (\ref{eq-condition_averaging_time}).
Moreover, Eq.~(\ref{eq-time_evolution_of_K}) reads for an arbitrary lattice
\begin{eqnarray} \label{eq-time_evolution_of_K_z_arbitarary_lattice}
 & & \frac{1}{L} \Kk_{AB} (t)   =   4 \alphaA \alphaB  z \lambdaA^2 \lambdaB^2 \Big[ 4 z -2 - 4 (z-1) \cos ( t V) \cos ( t U_B) - 2
   \cos (2 t V) \cos ( t U_B) \Big]
 \\ \nonumber
 &  & \hspace*{11cm} \text{ if } t \ll U_A^{-1}, J_{B}^{-1}, V   U_B^{-2}\;.
 \end{eqnarray}
%



\end{document}